\documentclass[twocolumn]{aastex62}
\usepackage{times}
\usepackage{epsfig,graphics}
\usepackage{amsmath, amssymb, graphicx, multirow}
\usepackage{textcomp}
\usepackage{natbib}
\usepackage{float}
\usepackage{color}
\usepackage{epstopdf}
\usepackage[countmax]{subfloat}
\usepackage{booktabs}
\usepackage[caption=false]{subfig}
\usepackage{mwe}
\usepackage[graphicx]{realboxes}
\usepackage{rotating}
\usepackage{longtable}
\newsubfloat{figure}
\shortauthors{Kalita et al.}

\accepted{ApJ}
\shorttitle{Optical variability of BL Lac}
\shortauthors{Kalita et al. 2022}
\begin{document}
\title{\Large {\bf Optical flux and spectral variability of BL Lacertae during its historical high outburst in 2020}}

\correspondingauthor{Nibedita Kalita}
\email{nibeditaklt1@gmail.com}

\author[0000-0002-9323-4150]{Nibedita Kalita}\thanks{Asian Forum for Polar Sciences Fellow}
\affiliation{Key Laboratory for Research in Galaxies and Cosmology, Shanghai Astronomical Observatory, Chinese~Academy of Sciences, 80 Nandan Road, Shanghai 200030, China}
\affiliation{Key Laboratory for Polar Science, MNR, Polar Research Institute of China, 451 Jinqiao Road, Shanghai 200135, China}

\author{Yuhai Yuan}
\affiliation{Center for Astrophysics, Guangzhou University, Guangzhou 510006, China}
\affiliation{Astronomy Science and Technology Research Laboratory of Department of Education of Guangdong Province, Guangzhou 510006, China}

\author{Minfeng Gu }
\affiliation{Key Laboratory for Research in Galaxies and Cosmology, Shanghai Astronomical Observatory, Chinese~Academy of Sciences, 80 Nandan Road, Shanghai 200030, China}

\author{Junhui Fan}
\affiliation{Center for Astrophysics, Guangzhou University, Guangzhou 510006, China}
\affiliation{Astronomy Science and Technology Research Laboratory of Department of Education of Guangdong Province, Guangzhou 510006, China}

\author{Yosuke Mizuno}
\affiliation{Tsung-Dao Lee Institute, Shanghai Jiao Tong University, 520 Shengrong Road, Shanghai, 201210, China}
\affiliation{School of Physics and Astronomy, Shanghai Jiao Tong University, 800 Dongchuan Road, Shanghai, 200240, China}

\author{Peng Jiang}
\affiliation{Key Laboratory for Polar Science, MNR, Polar Research Institute of China, 451 Jinqiao Road, Shanghai 200135, China}

\author{Alok C. Gupta}
\affiliation{Aryabhatta Research Institute of Observational Sciences (ARIES), Manora Peak, Nainital, 263001, India}
\affiliation{Key Laboratory for Research in Galaxies and Cosmology, Shanghai Astronomical Observatory, Chinese~Academy of Sciences, 80 Nandan Road, Shanghai 200030, China}

\author{Hongyan Zhou}
\affiliation{Key Laboratory for Polar Science, MNR, Polar Research Institute of China, 451 Jinqiao Road, Shanghai 200135, China}

\author{Xiang Pan}
\affiliation{Key Laboratory for Polar Science, MNR, Polar Research Institute of China, 451 Jinqiao Road, Shanghai 200135, China}

\author{Anton A. Strigachev}
\affiliation{Institute of Astronomy and National Astronomical Observatory, Bulgarian Academy of Sciences, 72 Tsarigradsko shosse Blvd., 1784 Sofia, Bulgaria}

\author{Rumen S. Bachev}
\affiliation{Institute of Astronomy and National Astronomical Observatory, Bulgarian Academy of Sciences, 72 Tsarigradsko shosse Blvd., 1784 Sofia, Bulgaria}

\author{Lang Cui}
\affiliation{Xinjiang Astronomical Observatory, Chinese Academy of Sciences, 150 Science 1-Street, 830011 Urumqi, China}

\begin{abstract}
BL Lacertae had undergone a series of historical high flux activity over a year, from August 2020 in the optical to VHE $\gamma$-rays. In this paper, we report on optical flux and spectral variability of the first historical maxima outburst event during October -- November in g, r and i bands with the 1.26m telescope at Xinglong observatory, China. We detected significant intranight variations with amplitude rising up to $\sim 30$\%, when the fastest variability timescale is found to be a few tens of minutes, giving an emitting region size of the order $10^{-3}$ pc, which corresponds to $\sim 100$ Schwarzschild radius of the central black hole, likely coming from some jet mini-structures. Unlike on intranight timescale, a clear frequency dependent pattern along with symmetric timescales ($\sim 11d$) of flux variation are detected on long timescale. The spectral evolution was predominated by flattening of the spectra with increasing brightness i.e., a bluer-when-brighter trend in 96\% of the cases. On the night before the outburst peak, the color indices clustered in two distinct branches in color--magnitude diagram within a period of $\sim$ 6 hours that is connected to a hard-soft-hard spectral evolution trend extracted from time-resolved spectra. Such trend has never seen in BL Lac or any other blazars before to the best of our knowledge. The results obtained in this study can be explained in the context of shock induced particle acceleration or magnetic re-connection in the jet where turbulent processes most likely resulted the asymmetric flux variation on nightly timescale.
\end{abstract}

\section{{\bf Introduction}}
BL Lacertae is the prototype of BL Lac objects which belong to the most energetic radio-loud class of active galactic nuclei (AGNs) known as blazars. The source is hosted by an elliptical galaxy of brightness R=15.5 located at a redshift of $z$ = 0.0668 in the northern hemisphere. Based on the location of synchrotron peak frequency in the spectral energy distribution (SED), it is classified as either a LBL (Low synchrotron peaked) or IBL (intermediate synchrotron peaked) blazar as it has been found to shift its peak energy on different occasions \citep{2011ApJ...743..171A, 2016ApJS..226...20F, 2018A&A...620A.185N}. The source has been detected in TeV energies \citep{2001ARep...45..249N, 2007ApJ...666L..17A} and found to show rapid flux variation within a few minutes in the $\gamma$-rays during high activity states which coincides with emergence of a new superluminal component from the radio core accompanied by changes in the optical polarization angle \citep{2013ApJ...762...92A}.

BL Lacertae (hereafter BL Lac) is one of the most frequently and well studied blazar in the multi-wavelength domain. The blazar has been a target of numerous multi-wavelength observing campaigns \citep[e.g.,][]{2002A&A...385...55H, 2003ApJ...596..847B, 2004A&A...424..497V, 2006A&A...456..105B, 2009A&A...507..769R, 2010A&A...524A..43R, 2019A&A...623A.175M}. The broadband spectrum of the source could be interpreted via either a single-zone or a two-zone SSC model, however an EC component + SSC model is the most likely explanation for the observed variability in the source \citep{2011ApJ...730..101A, 2021arXiv210812232S}. The optical spectra of the source shows presence of both broad and narrow emission lines during a not unusually faint state when the continuum polarization was estimated relatively low \citep{1996MNRAS.281..737C}. The study found that the H$_{\alpha}$ emission could be powered by thermal radiation from an accretion disc without significantly affecting the shape or polarization of the optical continuum. Later, \citet{2010A&A...516A..59C} found that the flux variation of H$_{\alpha}$ and H$_{\beta}$ emission lines was resulted by addition of gas in the broad line region (BLR). 

A study carried out by \citet{2008Natur.452..966M} found that a bright feature in the jet causes a multi-wavelength double flare originated in the acceleration and collimation zone in a helical magnetic field. Existence of helical magnetic field in BL Lac resulted in observed alternation of enhanced and suppressed optical activity that accompanied by hard and soft radio events, respectively \citep{2009A&A...501..455V, 2015ApJ...803....3C}. Evidence of multiple standing shocks along with helical magnetic fields was also reported from polarimetric space VLBI observations by RadioAstron \citep{2016ApJ...817...96G}.

Variation in the blazar flux over a time-scale of few minutes to less than a day is commonly known as intra-day variability \citep[IDV;][]{1995ARA&A..33..163W}, while Variability time-scales of weeks to a few months and months to years are known as short-term variability (STV) and long-term variability (LTV), respectively \citep{2004A&A...422..505G}. BL Lac is well known for its optical flux and polarization variability on diverse timescales and hence, has been observed by several observatories on different occasions and studies have been carried out to understand the physical properties \citep{1998MNRAS.299...47M, 2013MNRAS.436.1530R, 2015MNRAS.452.4263G, 2020ApJ...900..137W}. \citet{2020ApJ...900..137W} reported that turbulent plasma is responsible for multi-wavelength timescales of variability, lags in cross-frequency variation, and polarization properties where shock in the jet energizes the plasma which subsequently loses energy via synchrotron and inverse Compton radiation in a strong B-field of strength $\sim 3$G. It has found to show anti-correlated flux and polarization where PA was almost non-variable \citep{2014ApJ...781L...4G}. In several occasions micro-variations have been detected in the source accompanied with flattening of the spectra with increasing brightness popularly known as bluer when brighter trend \citep{1989Natur.337..627M, 2003A&A...397..565P, 2006A&A...450...39G, 2015MNRAS.450..541A, 2018Galax...6....2B}. Such micro-variations could be resulted by perturbations of different regions in the jet which cause localized injections of relativistic particles on time scales much shorter than the average sampling interval of the light curves where the cooling and light crossing time scales control the variations \citep{2003A&A...397..565P}. Lag between spectral and flux changes detected in long term study made with {\it WEBT} observations was explained in terms of Doppler factor variations due to changes in the viewing angle of a curved and inhomogeneous emitting jet by \citet{2007A&A...470..857P}.

Recently, the source has undergone a prolonged episodes of historical high flux activity starting from August 2020 till  August 2021, in the optical to VHE $\gamma$-rays ($>$ 100 GeV) \citep[ATel\#][]{2020ATel13930....1G, 2020ATel13933....1C, 2020ATel13963....1B, 2020ATel13964....1O}. In the optical, the source reached its first historical high flux state in October 5, 2020 with a recorded R--band magnitude of 11.73 $\pm$ 0.01 by the Kanata telescope \citep[ATel\#][]{2020ATel14081....1S}. After the first flare, the source attained an even brighter phase with magnitudes below R=11.5 recorded by WEBT Collaboration on 17 January 2021 \citep[ATel\#][]{2021ATel14342....1D} and in July 31st, it reaches the brightest state ever by going down to 11.271$\pm$0.003 mag \citep[ATel\#][]{2021ATel14820....1K}.

Following the Astronomer's telegram on the first enhanced activity of the source in 2020, we monitored the source starting from October 1 to November 23, 2020 in g, r, and i filter bands with a 1.26m telescope located at Xinglong Observatory in China. We recorded the 1st optical flare along with its raising and decaying phase during the series of high activity events. In this paper, we have investigated the temporal and spectral behavior of the blazar in optical band and are presenting our first result. The paper is structured as follows. In Section 2, we give the information about our photometric observations and describe the data analysis techniques. In Section 3, we present the analysis techniques that we used to investigate variability. We present the results in section 4, and discussion and conclusion in Section 5.

\section{{\bf Observations and Data Reductions}}

\begin{table*}
\caption{The comparison stars of BL Lac}
\label{compare}
\centering
\begin{tabular}{ccccccccccccc}
\hline
\hline
Comparison stars	&  $m_V \pm \sigma_V$&&	$m_R \pm \sigma_R$ &&$m_I\pm \sigma_I$  && $m_g$ && $m_r$ && $m_i$& \\
 (1)    &  (2)	&&	(3)	&&	(4)	&&	 (5)  && (6) && (7) &\\
\hline\hline
B   &12.78$\pm$0.04	&&11.93$\pm$0.05	&&11.09$\pm$0.06	&&13.67	&&12.01	&&11.17	&\\
C   &14.19$\pm$0.03	&&13.69$\pm$0.03	&&13.23$\pm$0.04	&&14.76	&&13.74	&&13.27	&\\\hline
\end{tabular}\\
Column:(1) The standard stars of BL Lac are labeled as B and C; (2), (3), $\&$ (4) represent magnitudes \\ with standard deviation at $V$, $R$, and $I$ bands, respectively and (5), (6), and (7) represent magnitudes \\at $g$, $r$, and $i$ bands, respectively.
\end{table*}

The photometric observations were carried out with the 1.26-m National Astronomical Observatory-Guangzhou University Infrared/Optical Telescope (NAGIOT) at Xinglong station of National Astronomical Observatories, Chinese Academy of Sciences (NAOC). This telescope is equipped with three SBIG STT-8300M cameras, whose CCD contains 3326$\times$2504 pixels and a view field of $6.0'\times 4.5'$. The system enables simultaneous photometry in three optical bands where the three filters adopt the standard SDSS $g$, $r$ and $i$ bands. The aperture radius used for aperture photometry was 1.2$\times$ FWHM, where FWHM is the average FWHM value of around ten bright stars in the same frame of BL Lac. For sky background, we selected a source free annulus region with inner and outer radius of 2.4$\times$ and 3.6$\times$ FWHM, respectively \citep[for detail see][]{2019RAA....19..142F}. The exact simultaneity of the observations at three bands is particularly suitable for studying the flux and color variations in blazars.

The observed images contain the bias, dark, flat-field and target images. We used 300s exposure time for each images in all the bands. The data reduction was carried out using the RAPP \citep[robust automated photometry pipeline,][]{2020PASP..132g5001H}, which includes the following steps. The observed images were corrected for bias, flat and dark current. Then RAPP automatically detects the position of the stars in each image, and it matches the images based on the position information of the stars. These images were used to create an overlay image which was later used to register the position of the stars in each CCD image, and the position of a star in different images are obtained. Finally, we carried out the aperture photometry process using the APPHOT package in IRAF\footnote{IRAF is distributed by the National Optical Astronomy Observatories, which are operated by the Association of Universities for Research in Astronomy, Inc., under cooperative agreement with the National Science Foundation.} (Image Reduction and Analysis Facility) software.	

\begin{figure}
\begin{center}
\includegraphics[angle=0,width=7.0cm]{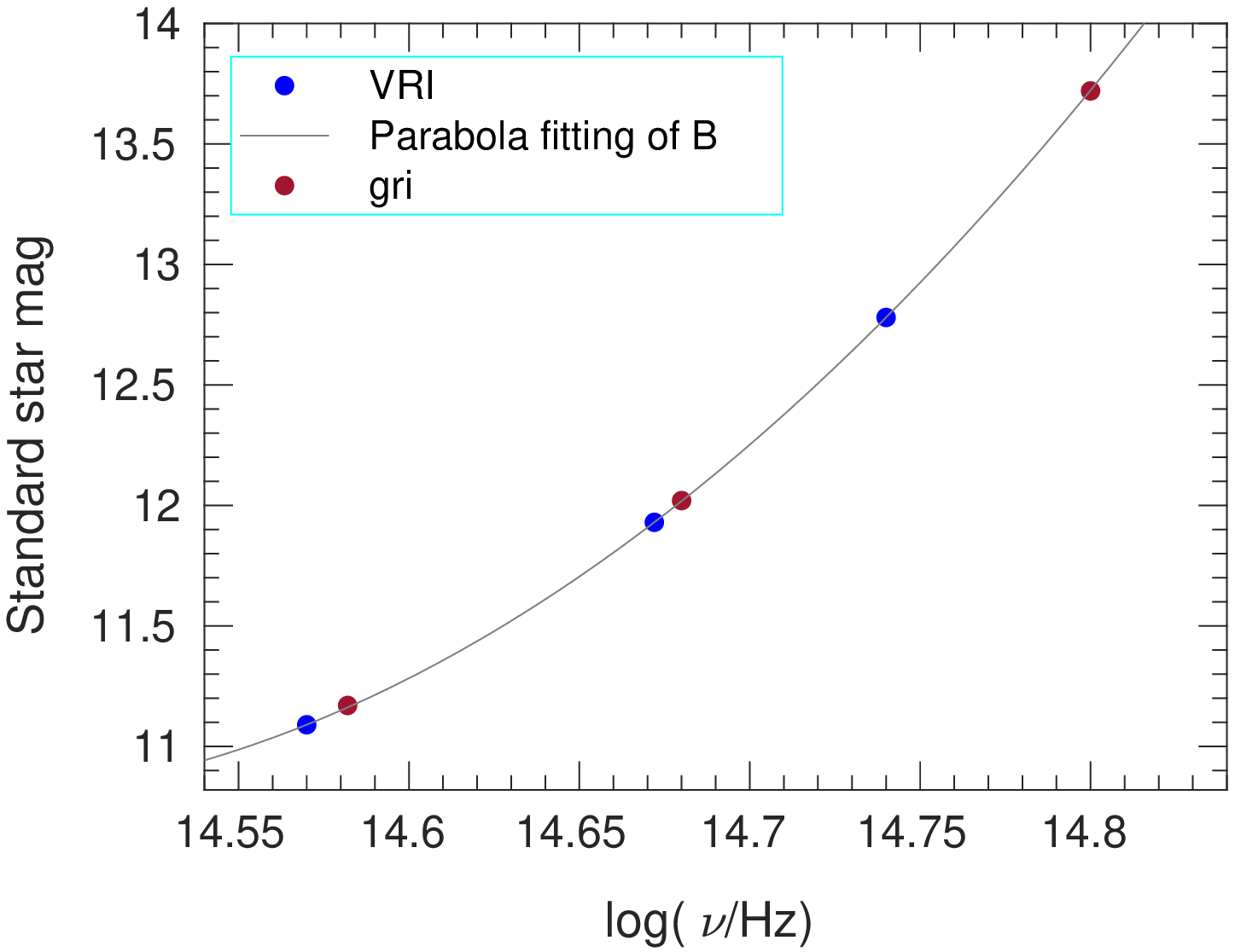}
\includegraphics[angle=0,width=6.7cm]{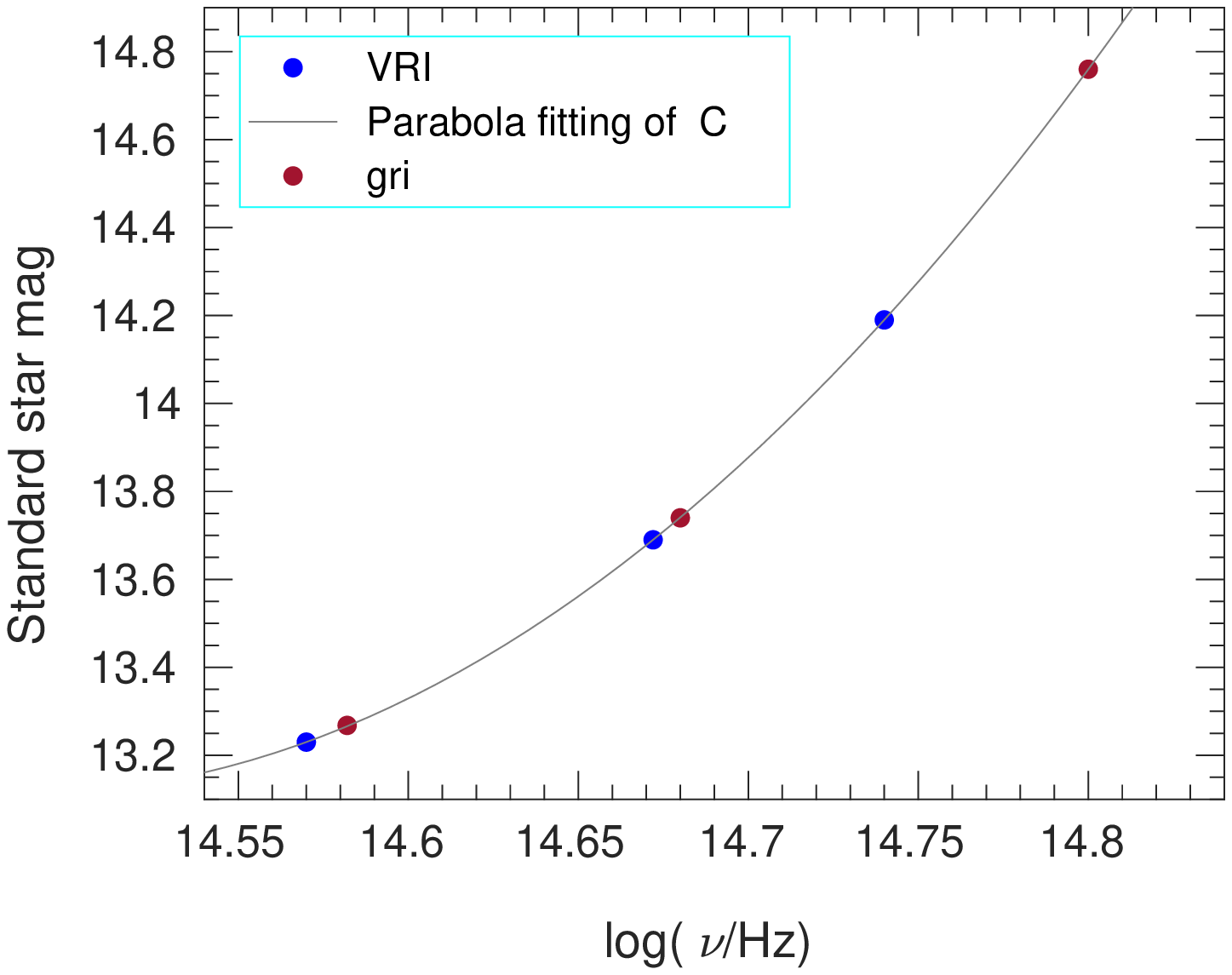}\\
\caption{The $gri$ fitting results of two comparison stars, B and C, of BL Lac. The black dots note $VRI$ magnitude, the red dots note $gri$ fitting results, and the solid line notes the polynomial fitting lines.}
\label{image:compare}
\end{center}
\end{figure}

We referred to \citet{1985AJ.....90.1184S} to obtain the standard stars for this source as listed in Table \ref{compare}. For all the comparison stars, based on $VRI$ magnitudes, we used a least square fitting method to obtain the $\textit{gri}$ magnitudes, $m_{\nu}=a\log^2\nu+b\log\nu+c$, here, $m_\nu$ is the magnitude at $\nu$-band ($\nu$= $V$, $R$ and $I$). Figure \ref{image:compare} gives the the fitting results, where, the black circles represent $VRI$ magnitudes, the red dots stand for $g$, $r$, and $i$ magnitudes, and the green curves show the least square fitting results. 

Following the above method, we extracted the instrumental magnitudes of the source and comparison stars. The light curves of the source during the observing period on daily and monthly timescales are shown in Figure 2 and 3, respectively. For calibrated source magnitude, we took average of the differences of the source and the two comparison stars, and the corresponding uncertainty was estimated using error propagation method.

\begin{figure*}
\label{fig1}
\centering
\begin{minipage}{0.4cm}
\rotatebox{90}{{\large {Magnitude}}}
\end{minipage}
\begin{minipage}{\dimexpr\linewidth-1.10cm\relax}
\hspace*{0.0cm}
\mbox{\subfloat{\includegraphics[scale=0.45]{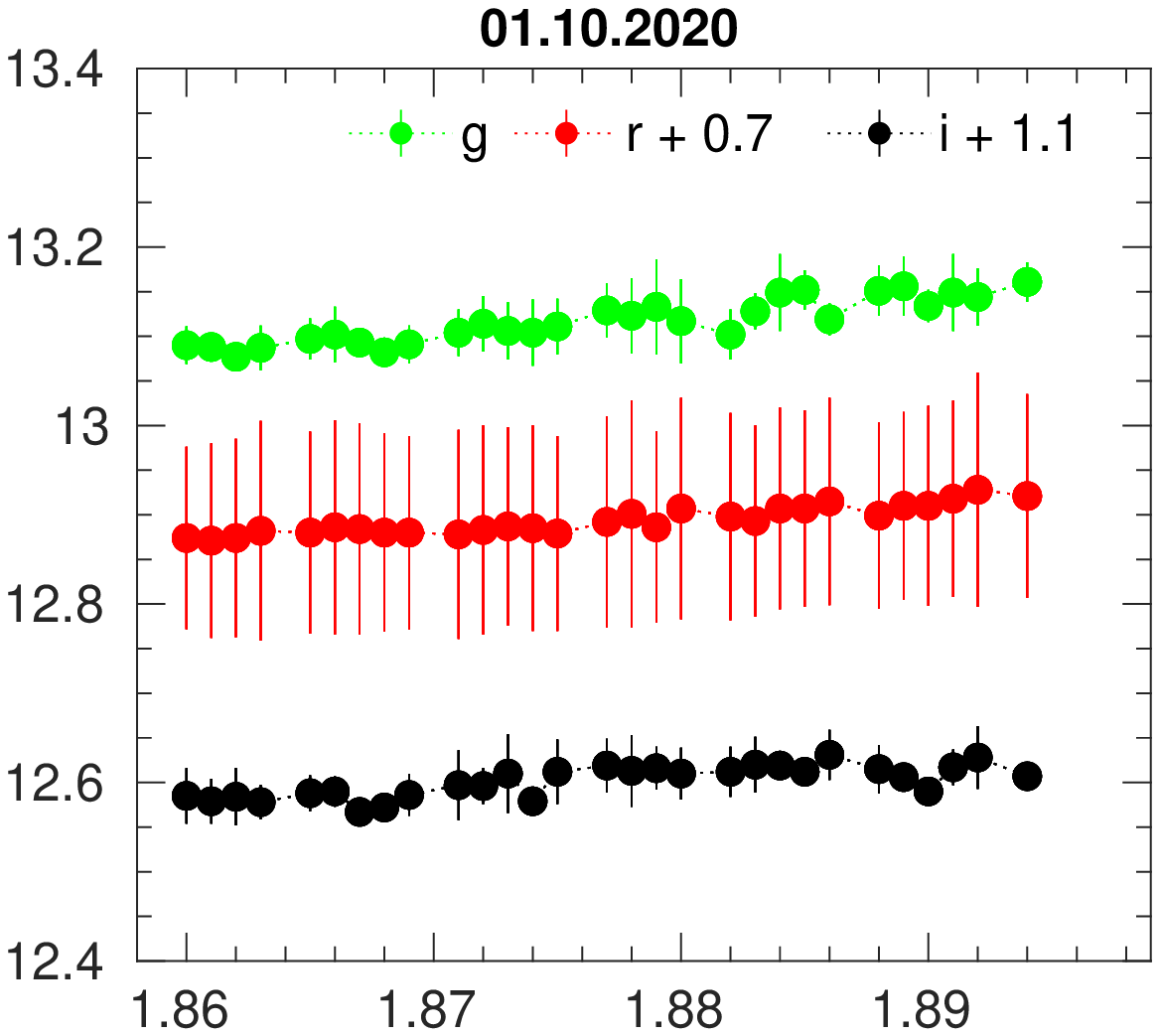} }\quad
\subfloat{\includegraphics[scale=0.45]{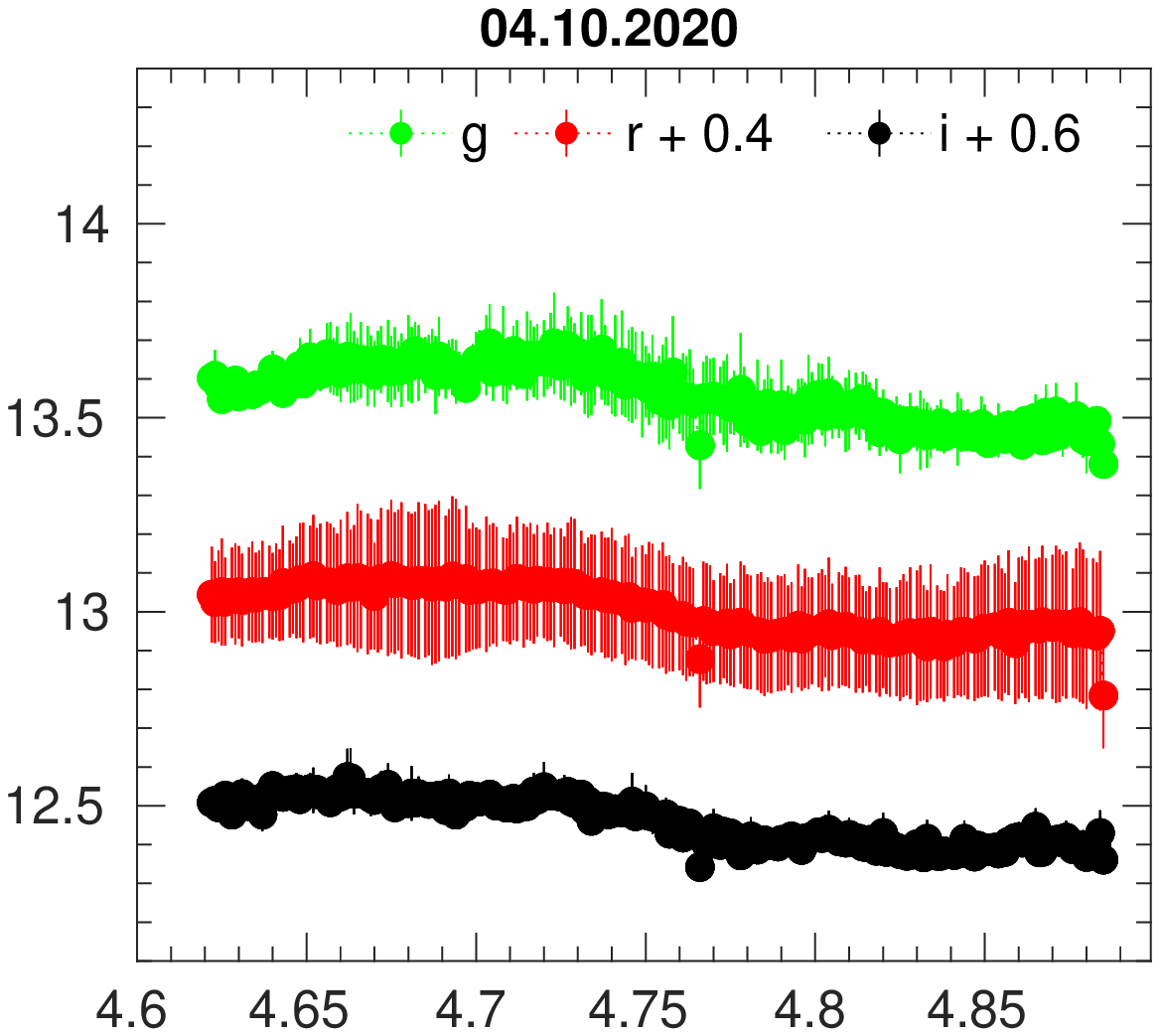} }\quad
\subfloat{\includegraphics[scale=0.45]{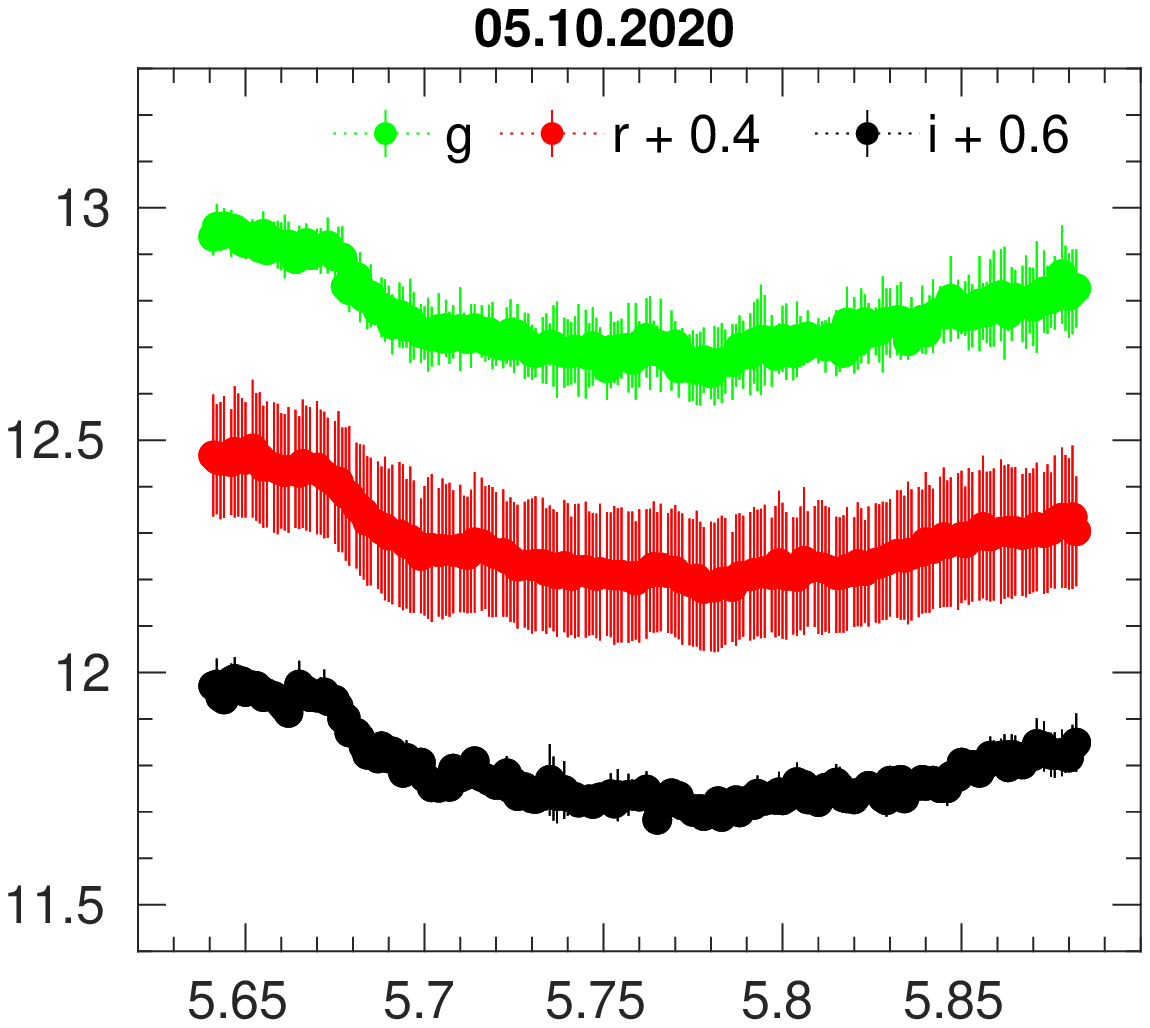}}}
\newline
\hspace*{0.0cm}
\mbox{\subfloat{\includegraphics[scale=0.45]{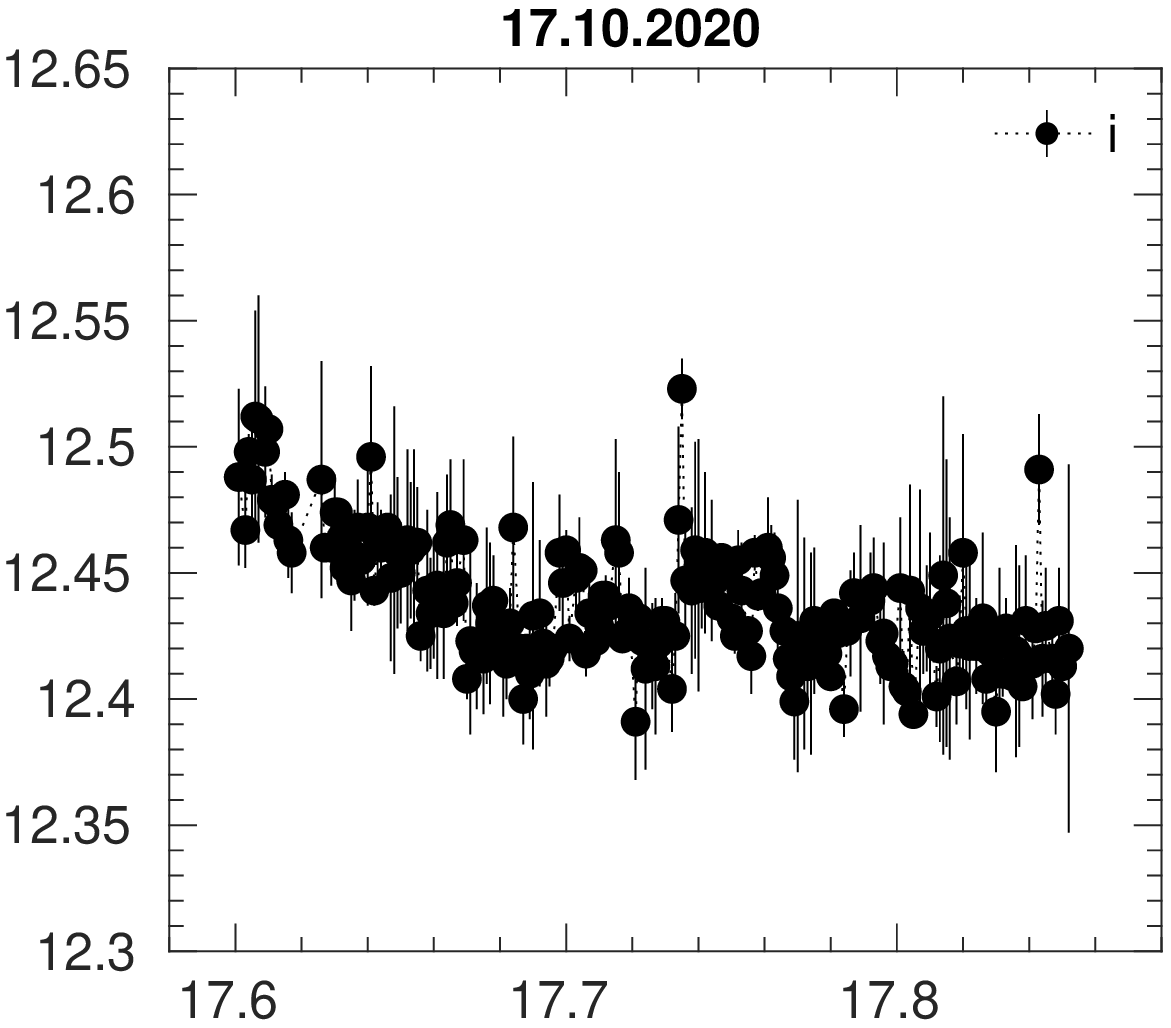} }\quad
\subfloat{\includegraphics[scale=0.45]{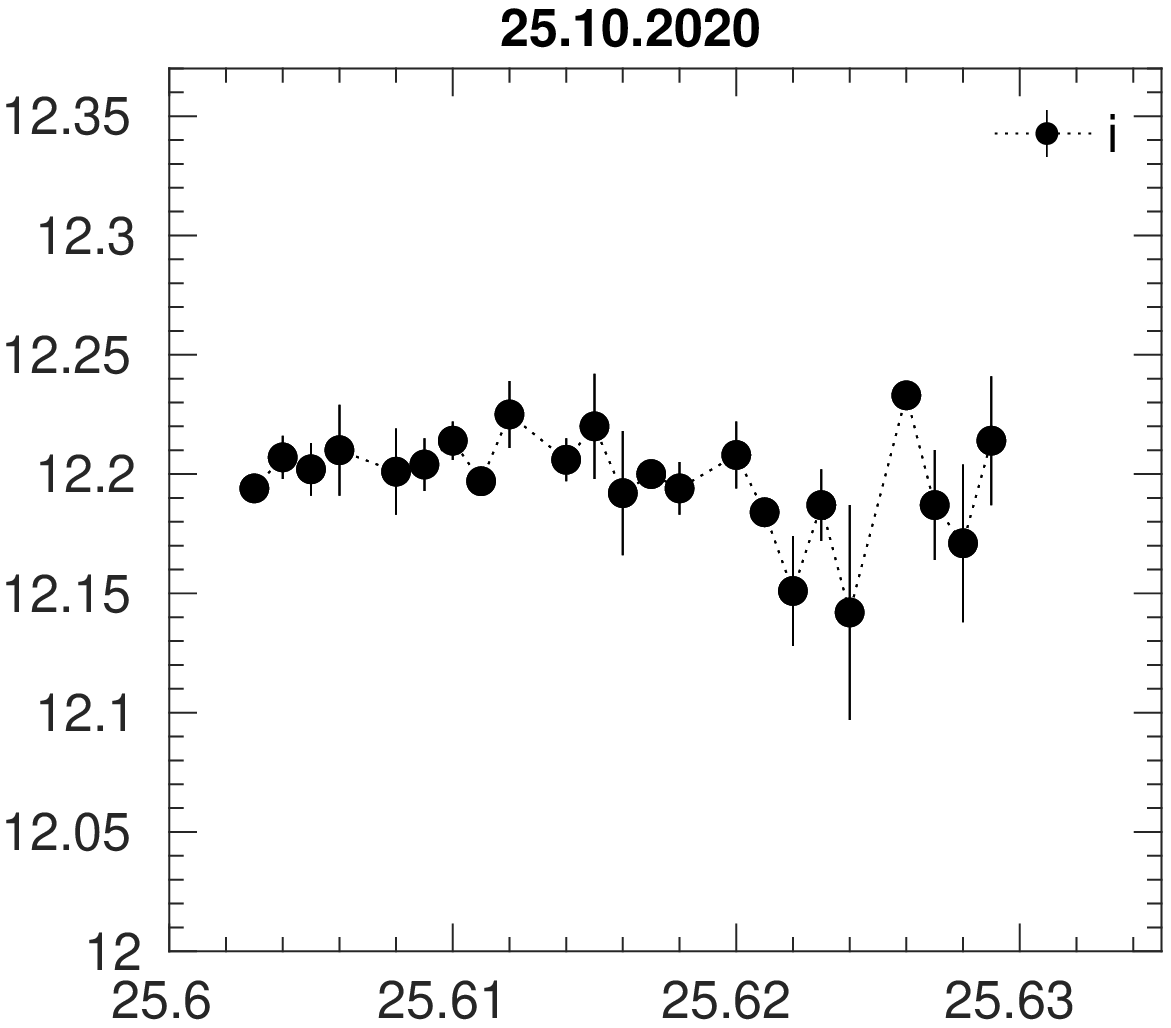} }\quad
\subfloat{\includegraphics[scale=0.45]{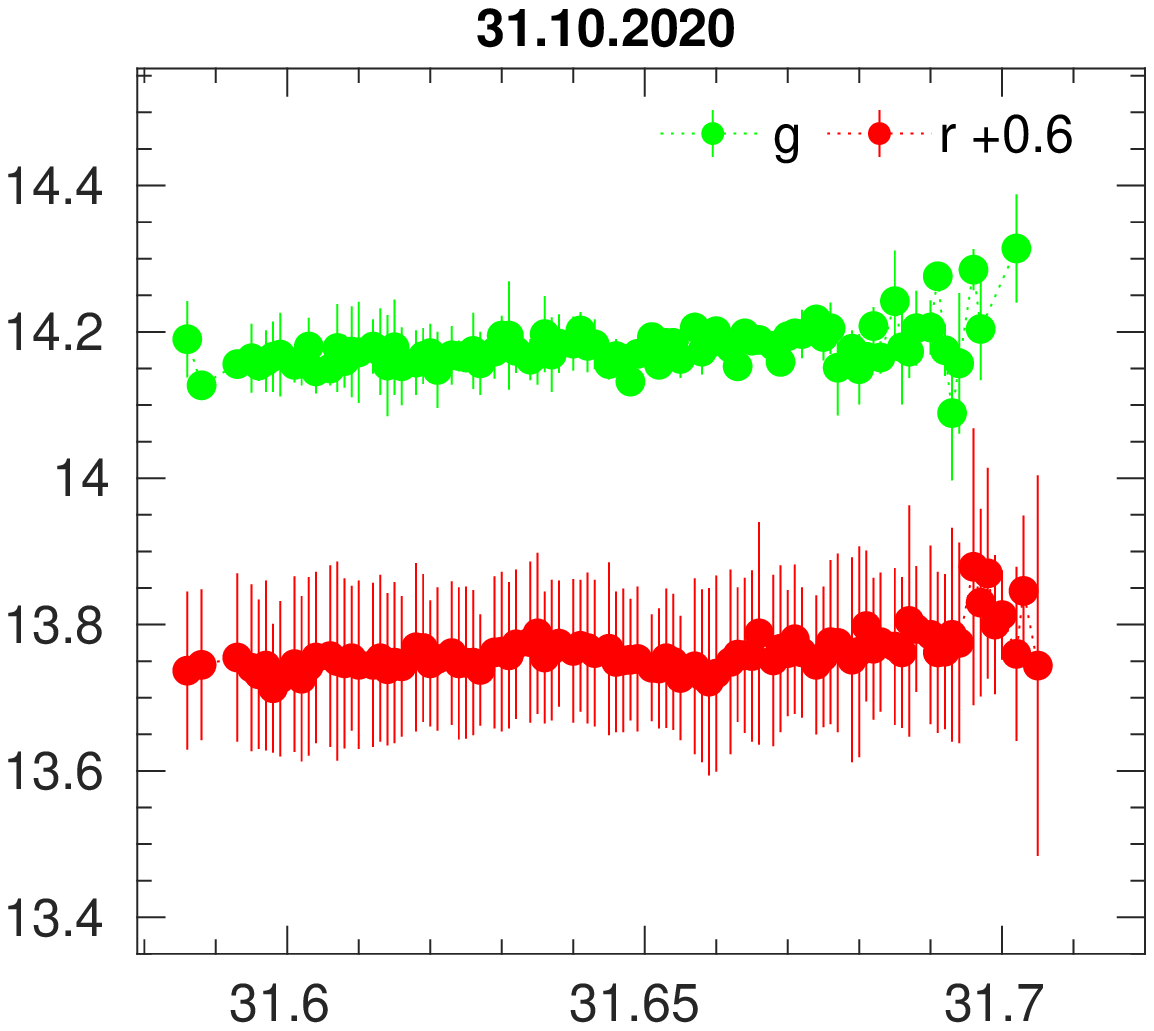} }}
\newline
\hspace*{0.0cm}
\mbox{\subfloat{\includegraphics[scale=0.45]{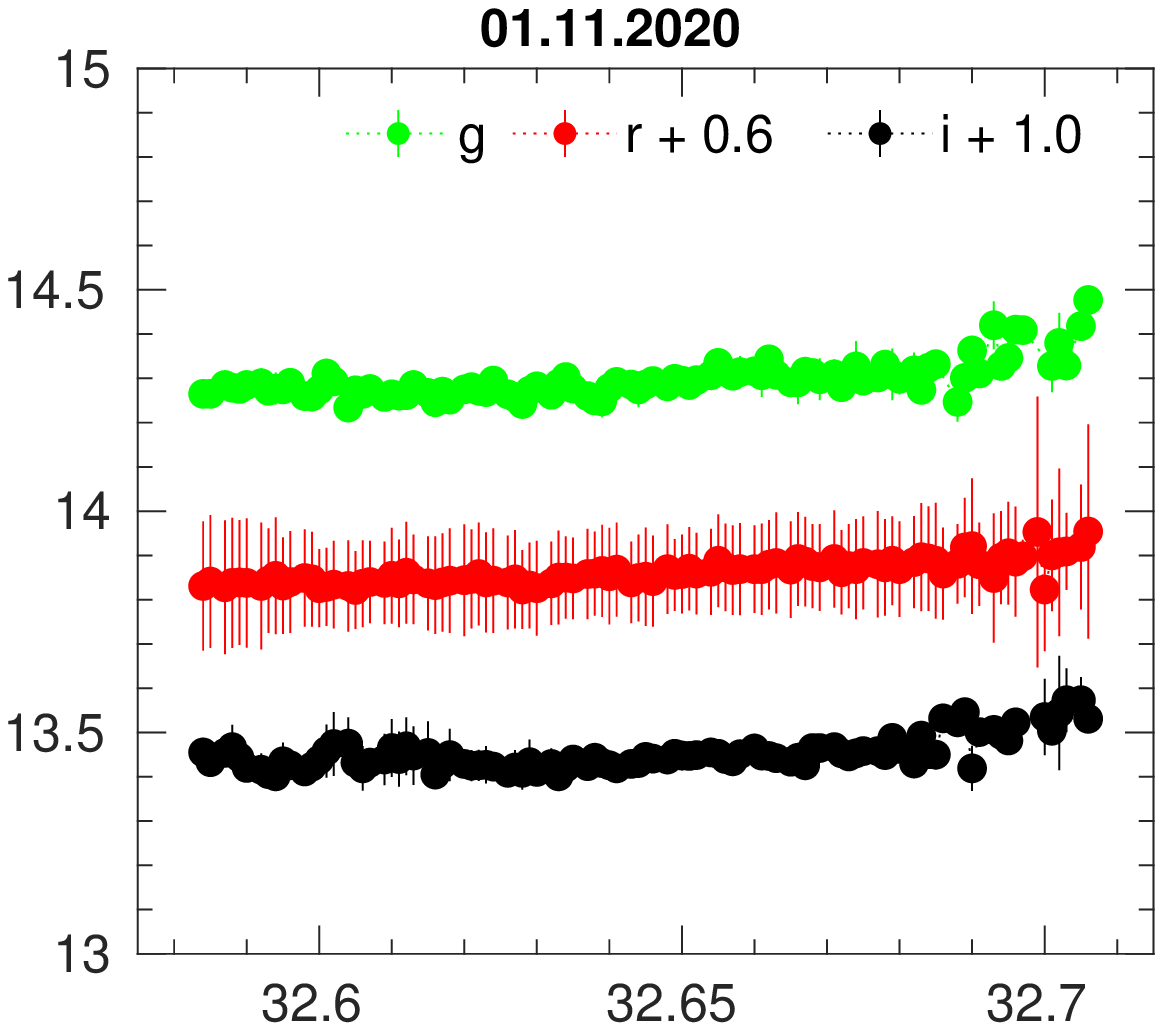} }\quad
\subfloat{\includegraphics[scale=0.45]{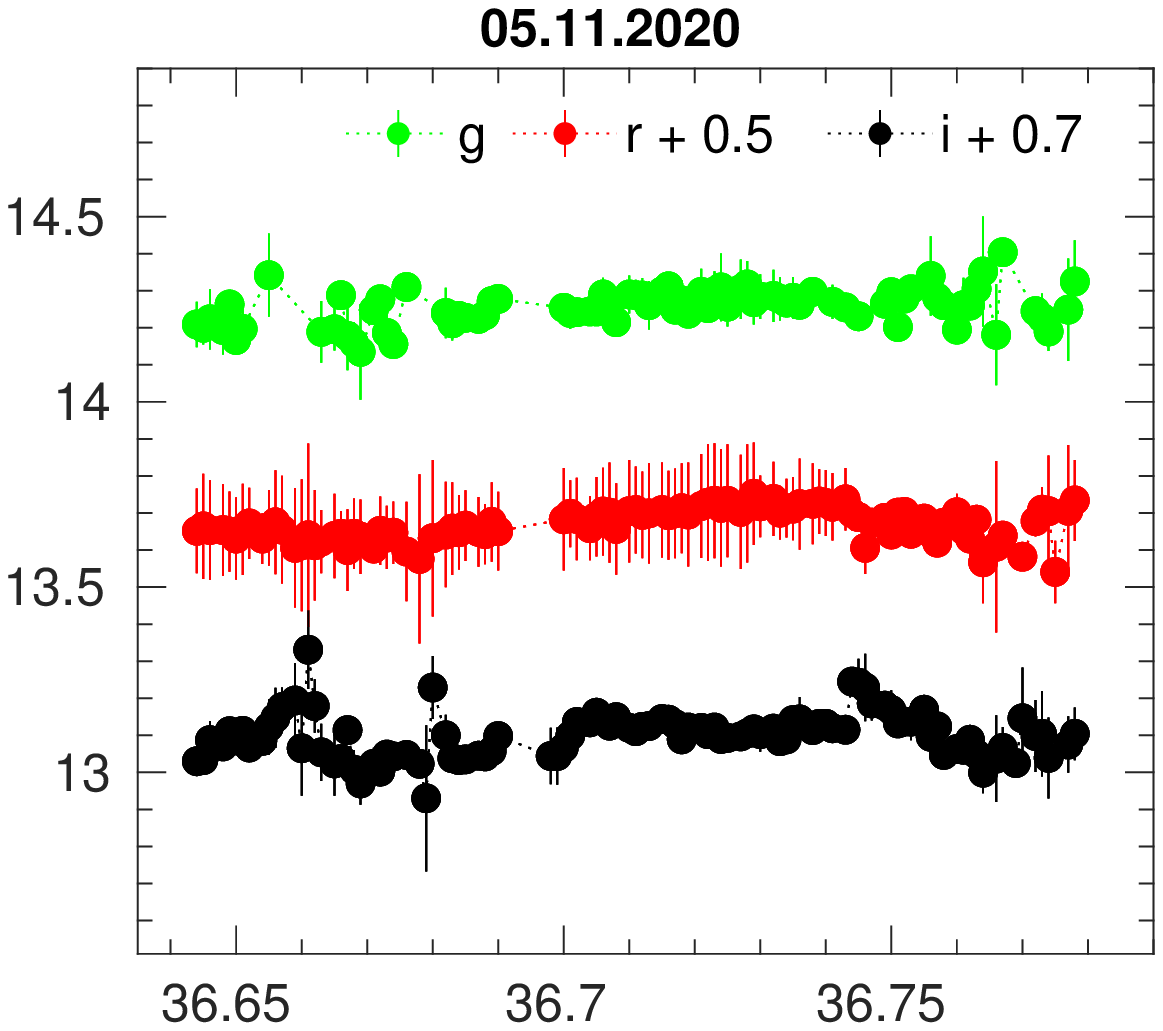} }\quad
\subfloat{\includegraphics[scale=0.45]{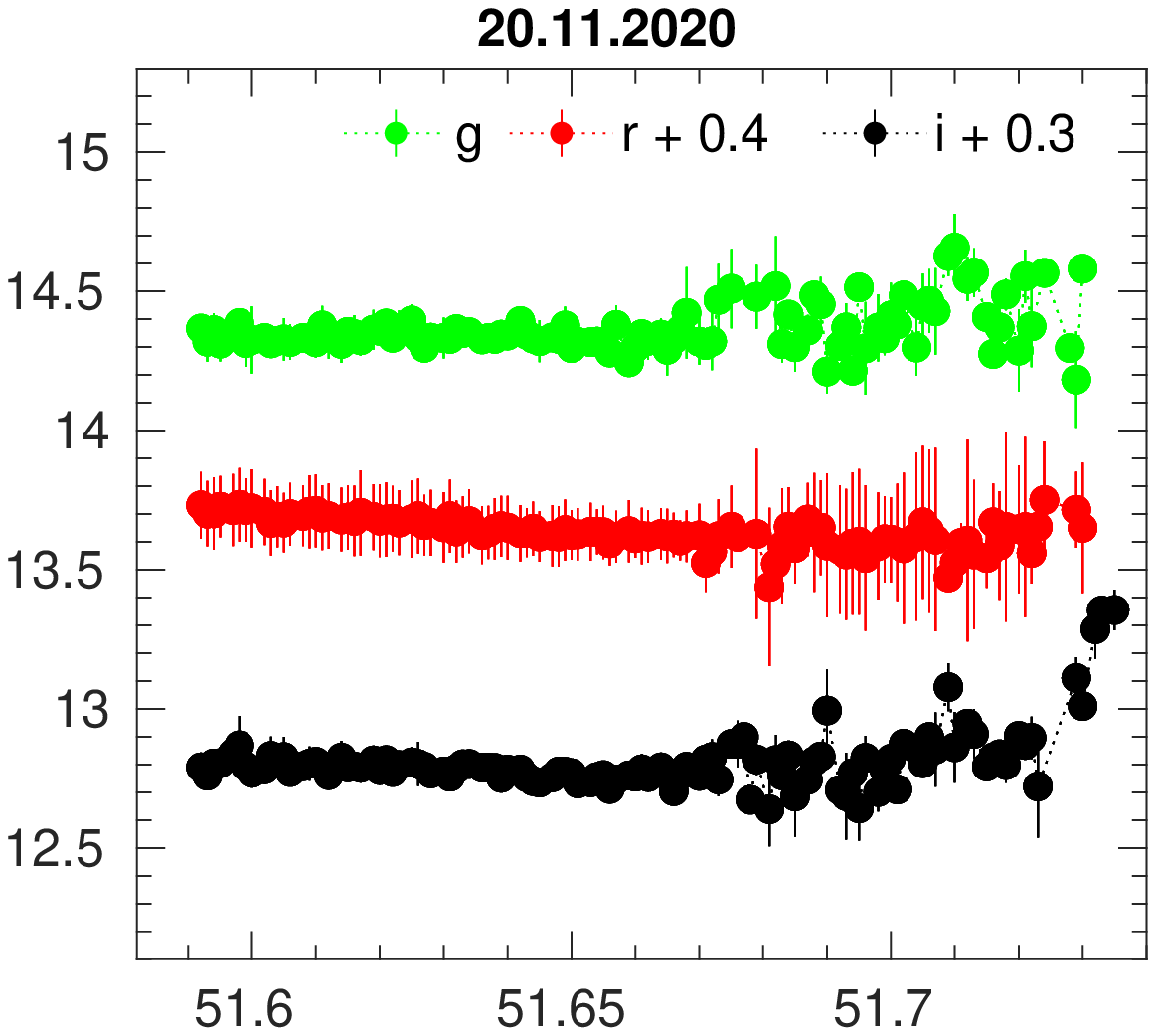} }}
\newline
\hspace*{5.9cm}
\mbox{\subfloat{\includegraphics[scale=0.45]{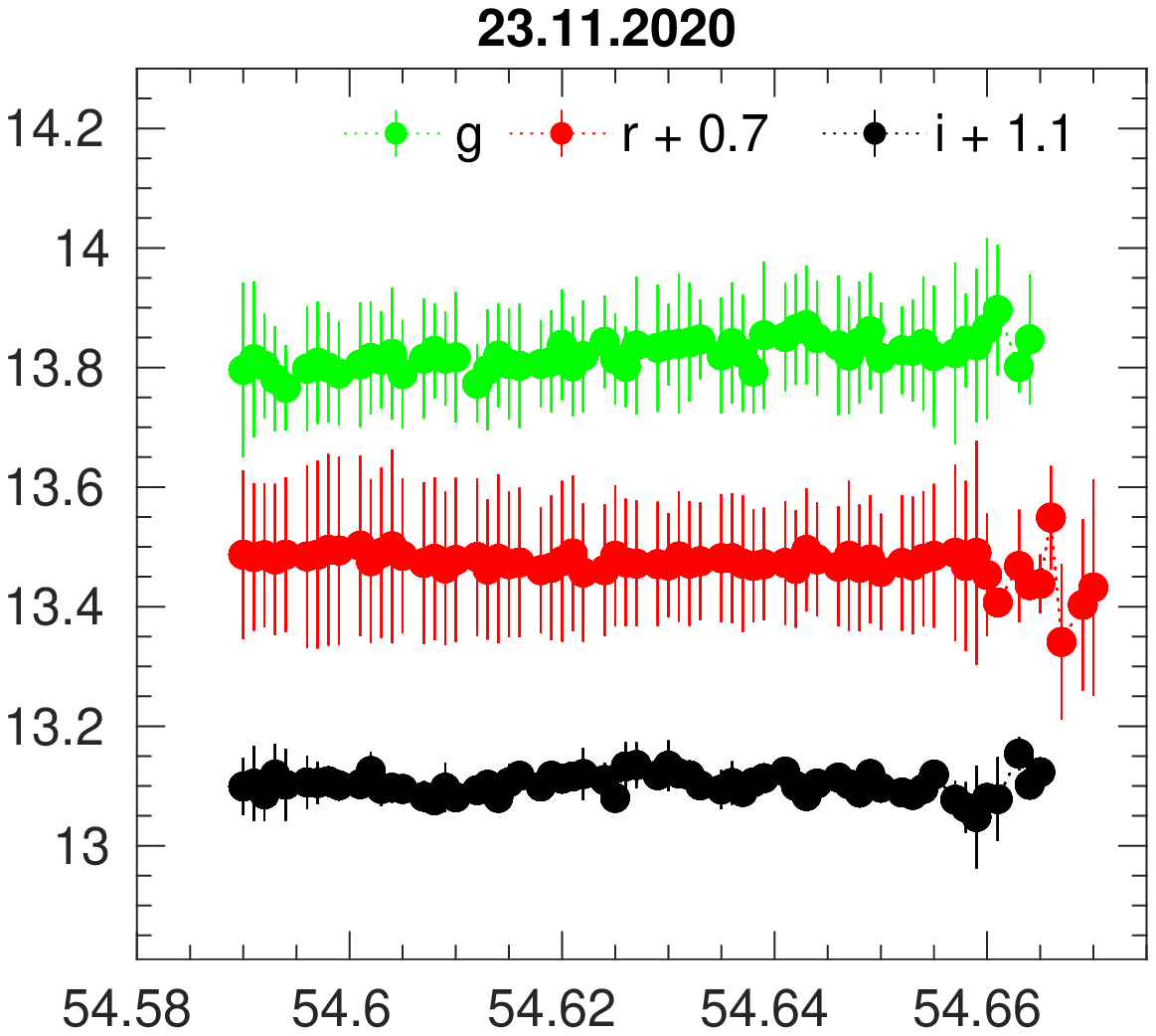} }}
\newline
\newline
\vspace*{0.6cm}
\hspace*{7.0cm}
{{\large{ Time (JD + 2459122)}}}
\caption{Nightly optical light curves of BL Lac in g, r, and i filter bands with 100s binning. In some nights only one (17th and 25th October) or two (31st October) filter bands observations are available. observing dates are shown at the top of each plot.}
\end{minipage}
\end{figure*}

\section{{\bf Statistical tests to study Variability}}
In order to detect micro-variation in time series observations of AGN, several statistical tests have been introduced and used. Among which the C-, F- and $\chi^{2}$-tests have been used widely. In addition to those, recently the Power-enhanced F-test and Nested ANOVA test have been gaining popularity due to their robustness to detect micro-variability precisely and accurately \citep[][see also \citep{2012MNRAS.420.3147G, 2021ApJS..253...10F, 2021ApJS..257...41K}]{2014AJ....148...93D, 2015AJ....150...44D}. In this study we used three methods; F-test, $\chi^{2}$-test and Nested ANOVA test to quantify variability of the source, which are briefly discussed below.

\subsection{F-test}
In the F-test \citep{2010AJ....139.1269D}, the source differential variance is compared to the differential variance of the comparison star (CS). The $F$ value is calculated as
\begin{equation}
F_1=\frac{Var(BL-CSA)}{Var(CSA-CSB)},
F_2=\frac{Var(BL-CSB)}{Var(CSA-CSB)},
\end{equation}

where Var(BL-CSA), Var(BL-CSB) and Var(CSA-CSB) are the variances of differential instrumental magnitudes of BL Lac and CS A, BL Lac and CS B, and CS A and CS B, respectively. An average of $F1$ and $F2$ gives the $F$ value which is compared with the critical $F$-value, $F^\alpha_{\nu_{bl},\nu_\ast}$, where $\nu_{bl}$ and $\nu_{\ast}$ are the number of degrees of freedom for the blazar and comparison star respectively, calculated as (N - 1) with N being the number of measurements, and $\alpha$ is the significance level set for the test which was 99\% (2.576$\sigma$) in this case. If the average $F$-value is larger than the critical value, the light curve is variable at a confidence level of 99 percent. \\

\subsection{$\chi^{2}$ -- test of variance}
In order to check the presence of variability
, we also performed a $\chi^{2}$-test. The null hypothesis is rejected when the statistic exceeds a critical value for a given significance level, $\alpha$. The statistic is given as,

\begin{equation}
\chi^{2} = \sum\limits_{i=1}^N {\frac { (V_{i} - \overline{V})^{2}}  {\sigma_{i}^{2}}  },
\end{equation}

where, $\overline{V}$ is the mean magnitude, and $V_{i}$ is the magnitude corresponds to the $i^{th}$ observation having a standard error $\sigma_{i}$. We took the average of $\chi^{2}$ values obtained from differential magnitudes related to the two CSs. This statistic is then compared with a critical value $\chi_{\alpha, \nu}^{2}$ where $\alpha$ is the significance level set same as in F-test and $\nu = N - 1$ is the degree of freedom. A smaller value of $\alpha$ assures more improbable that the result is produced by chance. Presence of variability is confirmed if $\chi^{2}$ $>$ $\chi_{\alpha, \nu}^{2}$. 

\subsection{Nested ANOVA test}
In nested ANOVA test \citep{2015AJ....150...44D}, multiple field stars are used as reference to estimate the blazar differential photometry without using any comparison stars. We used groups of replicated observations that compare the dispersion of the individual differential magnitudes of the source within the groups with the one between the groups \citep[discussed in detail in][]{2021ApJS..257...41K}.

Here, we used two reference stars, the comparison + reference stars used in the previous tests. We divided the time series observations into different temporal groups, $n$, where each group contain $m = 5$ observations, then we estimated the mean square due to groups ($MS_G$) and due to nested observations in groups ($MS_{O(G)}$) with dof $\nu_{1} = n - 1$ and $\nu_{2}=n(m - 1)$,
 respectively. The statistic is given by
\begin{equation}
F_{} = \frac{MS_{G}}{MS_{O(G)}},
\end{equation}

If the $F-$value exceeds the critical value $F_{\nu_{1}, \nu_{2}}^{(\alpha)}$ at a significance level of 99\% ($\alpha= 0.01$), the null hypothesis will be rejected. 

\subsection{Variability Amplitude}
To estimate the variability amplitude of the light curves (LCs), we use the variability amplitude
defined by \citet{1996A&A...305...42H} as follows
\begin{eqnarray}
A = \sqrt {(A_{max} - A_{min})^{2} - 2\sigma^{2}}
\end{eqnarray}

where A$_{max}$ and A$_{min}$ are the maximum and minimum magnitudes in the blazar LCs and $\sigma$ is the mean error.
\begin{figure}
\label{ltv}
\centering
\includegraphics[scale=0.58]{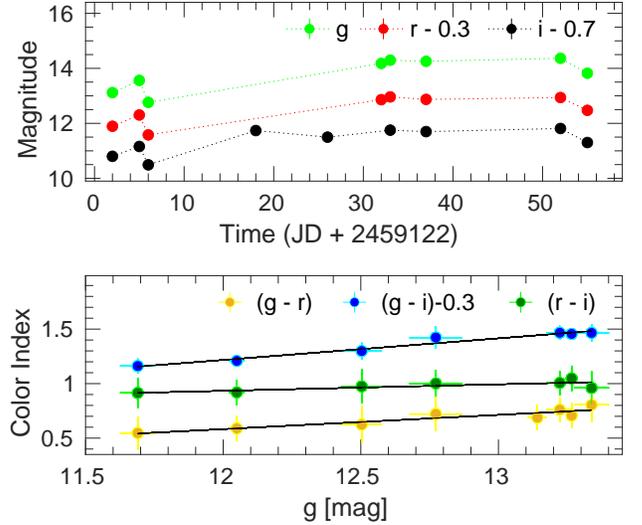}
\caption{Top: long timescale optical g-, r-, and i-band light curves over the whole observing period of BL Lac during October--November 2020. Bottom: color indices with respect to g-band brightness. The black lines represent a least squared fitting to the data points.}
\end{figure}

\begin{figure*}
\label{dcf}
\centering
\mbox{\subfloat{\includegraphics[scale=0.5]{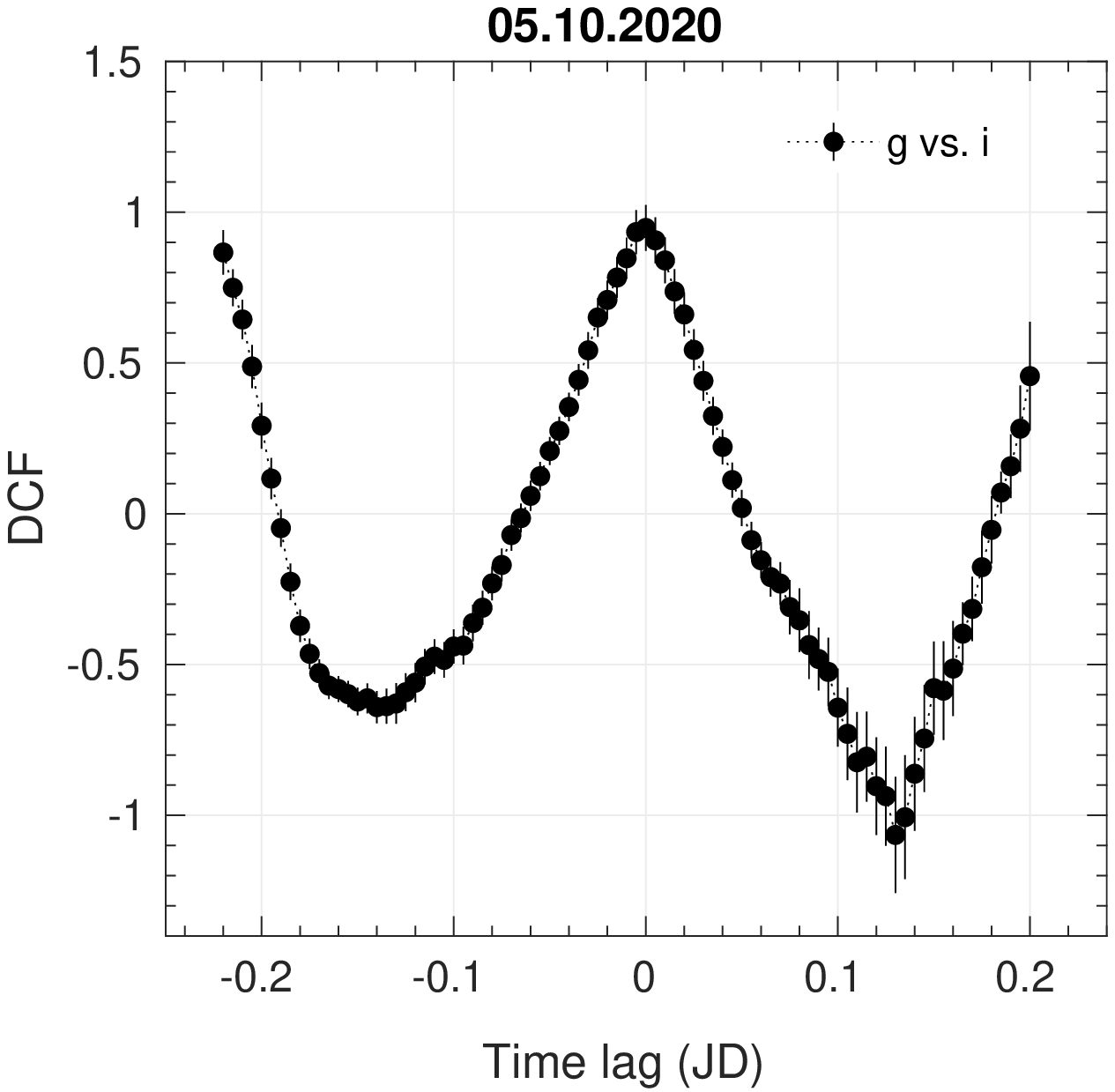} }\quad
\subfloat{\includegraphics[scale=0.52]{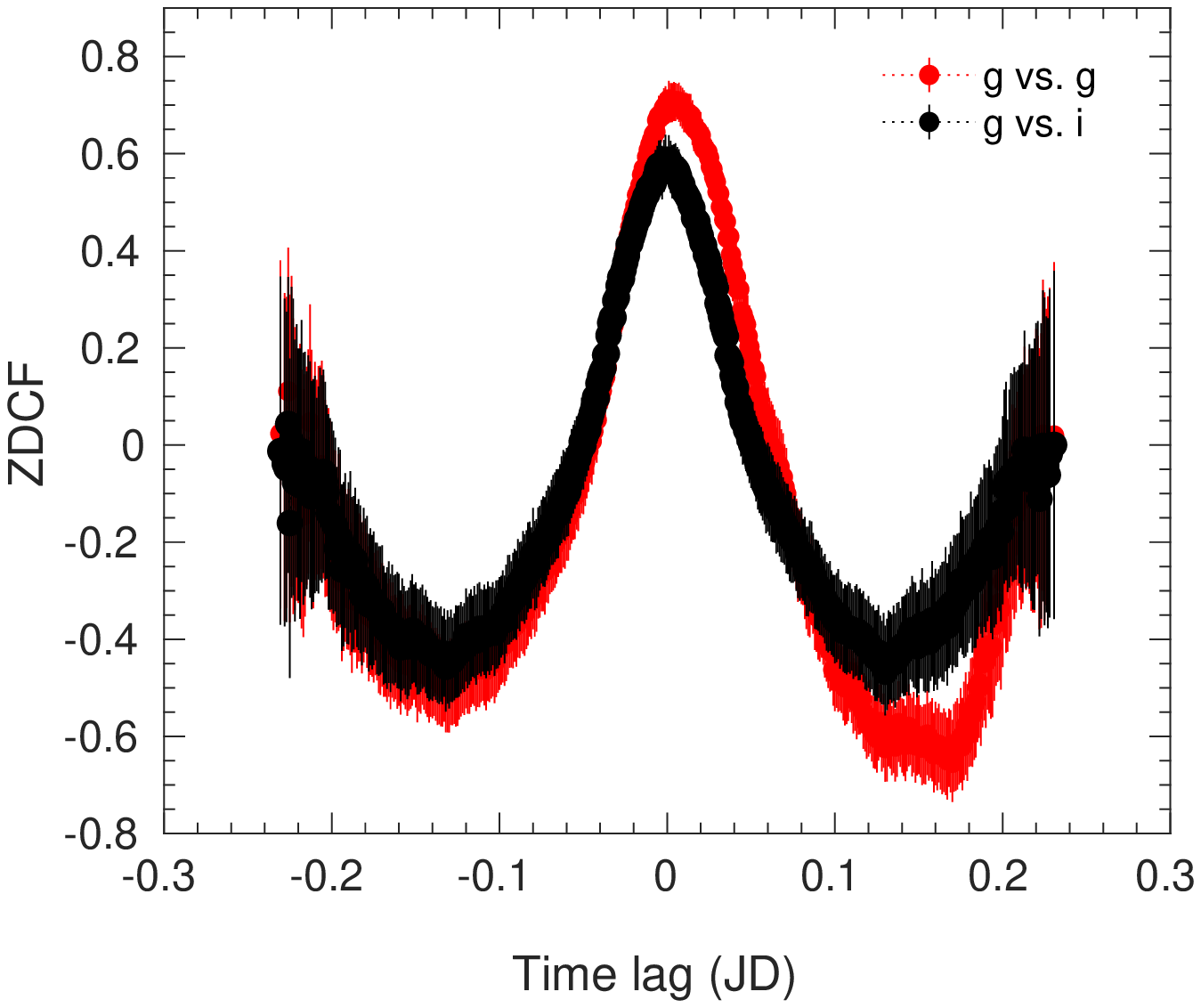} }}
\caption{An example of correlation between g and i bands using the DCF method is shown in 1st panel. Second panel shows the z-transformed DCF relation estimated using the code provided by \citet{1997ASSL..218..163A} where we used binning of 60, uniform sampling of the LC. Errors were estimated via MonteCarlo error approximation. The plot includes all the data points. Both tests show similar correlation pattern.}
\end{figure*}

\subsection{Discrete Correlation Function}
To check the correlation between the optical g, r, and i bands, we applied the Discrete Correlation Function (DCF; \cite{1988ApJ...333..646E}), which is one of the best methods to investigate a correlation between two unevenly sampled time-series data. A detailed description of the method we used in this work was given in \citet{2019ApJ...880...19K}. To measure the exact values of the peak and corresponding lag, we fitted the DCF peak by a Gaussian model of the following form: 

\begin{equation}
DCF(\tau)=a \times \exp \Bigl[\frac{-(\tau - m)^{2}}{2 \sigma^{2}}\Bigr]
\end{equation}
Here, $m$, $a$, and $\sigma$ represent the time lag at which DCF peaks, the peak value of the DCF, and the width of the Gaussian function, respectively. 

\section{ Results}
\subsection{ Flux and color variability}
We carried out the variability analysis of the light curves with the F-test, $\chi^{2}$-test, and nested-ANOVA test, which are discussed in the previous section. IDV results from these three analysis are presented in Table 2. The final remarks on the LCs in Table 2 were made based on F--statistic values and probability of rejecting the null hypothesis estimated from the three methods. If all three statistics are higher than the respective critical values ($F_{critical}$ and $\chi^{2}_{\alpha, \nu} $) at 99.9\% confidence level then we labeled the light curve as variable. If any two of the three statistic values are higher than the respective critical values at 99\% confidence level, then it was termed as probable variable (PV). Finally the non-variable (NV) one applies when the statistics do not satisfy the above mentioned conditions.

From our analysis, we detected significant intraday variability near and at the peak of the outburst which took place on October 5. During this phase the amplitude of variation reached up to $\sim$ 30\%. The significant detection of variability decreases towards the end of October. In November, the source showed mixed variability behavior. A significant variation was found on November 20, only in the i-band LC. In the observed period, out of 25 intra bands LCs, 10 are significantly variable, 4 are non-variable and 11 are probably variable. Interestingly, the highest amplitude of variation was found not at or near the outburst peak, but by the end of our monitoring program, on 20 November having a value $\sim$ 46\%. We do not see any frequency dependent variability from the amplitude analysis. 

The variability timescale for intraday flux is estimated using the following formula:

\begin{equation}
t_{v}=\Delta t \times \frac{ln2}{ln(f_{2}/f_{1})}
\end{equation}

where $f_{1}$ and $f_{2}$ are the flux values at time $t_{1}$ and $t_{2}$,
and $\Delta t$ is the difference between $t_{1}$ and $t_{2}$. We computed $t_{v}$ for all pairs of data points in the light curves and search for the shortest value in each. The shortest timescale of variability detected for each variable and probably variable LCs are given in Table 2. Since eq. 6 does not include the noise contribution in it thus, it is possible that it could detect some random outlying points that are close in time which could give a spurious small $t_{v}$ value. In order to avoid such error we incorporated the error values in equation 6 by two conditions; we considered all possible pairs of flux values that satisfy the conditions $f_{2}$ $>$ $f_{1}$ and $f_{2}$ $-$ $f_{1}$ $>$ 3($\sigma_{f_{1}}$ + $\sigma_{f_{2}}$)$\slash$2, where $\sigma_{f_{1}}$ and $\sigma_{f_{2}}$ are the uncertainties corresponding to the flux measurements $f_{1}$ and $f_{1}$ , respectively \citep{2013ApJ...773..147J}. Most of the r-band LCs could not pass the second condition due to comparatively high error bars in the this filter. Inclusion of errors change the values a little bit and these are listed in the last column of table 2. On intraday timescale, we detected a variety of variability timescales ranging from $\sim$ 34 hours to 52 minutes considering both probably variable and variable LCs with observation period $\sim$ 3 hours. If we consider only the variable instances from second estimation, the minimum variability timescale is $1^{h}$ $18^{m}$. Here again, the shortest variability timescale correspond not to the outburst peak but to the decaying phase near the end of the monitoring program (see Table 2).

\begin{sidewaystable*}
\vspace{8cm}
\begin{longtable}{ccccrcrrcrcrccccccccc}
\caption{\bf Observing log and results of variability analysis}
\\

\hline\hline \multicolumn{1}{c}{\textbf{Date of Obs.}} &\multicolumn{1}{c}{\textbf{Filter}}    &\multicolumn{1}{c}{\textbf{Obs.}}   &&\multicolumn{2}{c}{\textbf{{\it F}-- test}}   &&&\multicolumn{2}{c}{\textbf{$\chi^{2}$-- test}} &&&&\multicolumn{3}{c}{\textbf{Nested {\it ANOVA} test}}  &&\multicolumn{1}{c}{\textbf{FS}} & \multicolumn{1}{c}{\textbf{A}} & \multicolumn{1}{c}{\textbf{$t_{v}$}} &\multicolumn{1}{c}{\textbf{$t_{ve}$}}\\

                               \cline{4-11} \cline{14-16}

\multicolumn{1}{c}{\textbf{(dd.mm.yyyy)}}&\multicolumn{1}{c}{\textbf{band}} &\multicolumn{1}{c}{\textbf{duration}} &\multicolumn{1}{c}{\textbf{d($\nu_{1}$, $\nu_{2}$)}} &\multicolumn{1}{c}{\textbf{$F$}}    &&\multicolumn{1}{c}{\textbf{ $F_{c}$}}&&\multicolumn{1}{c}{\textbf{ $\chi^{2}$}} &&\multicolumn{1}{c}{\textbf{ $\chi_{\alpha, \nu}^{2}$}}     &&&\multicolumn{1}{c}{\textbf{dof($\nu_{1}$, $\nu_{2}$)}}    &\multicolumn{1}{c}{\textbf{$F$}}  &\multicolumn{1}{c}{\textbf{$F_{c}$}}   &&\multicolumn{1}{c}{\textbf{}} & \multicolumn{1}{c}{\textbf{\%}}& \multicolumn{1}{c}{\textbf{}} & \multicolumn{1}{c}{\textbf{}}\\

&\multicolumn{1}{c}{} & &(1) &(2)& &(3)&&(4)&&(5) &&&(6) &(7)&(8) &&(9)& (10)& (11)& (12) \\\hline
\endfirsthead
\endhead
\endfoot

\hline \multicolumn{21}{l}{{   {\bf Column:}(1) degrees of freedom (dof) in the F-statistic and $\chi^{2}$- distribution. (dof+1) represents no. of data points obtained during each night; (2) F value for F-test; }} \\
\multicolumn{21}{l}{{(3) \& (8) critical values at 99.9$\%$ ($F_{c}$);  (6) dof in the numerator and the denominator in ANOVA-statistics; (7) F value for nested-ANOVA test; (9) final variability status}} \\
\multicolumn{21}{l}{{ (V = variable; NV = non--variable; PV = probable variable); (10) amplitude of variation; (11) Timescale of variability;  (12) $t_{v}$ estimated incorporating errors.}} \\

\endlastfoot
01.10.2020&g &1$^{h}$0$^{m}$  &28  &2.84   &&2.44 &&263.46    &&48.28   && &6, 21    &17.45    &7.40    &&V  & 7.42 & 6$^{h}$37$^{m}$& 76$^{h}$00$^{m}$\\
          &r &,,              &28  &3.01   &&2.44 &&626.62    &&,,      && &6, 21    &19.06    &7.40    &&V  & 15.04& 9$^{h}$40$^{m}$& NA\\
          &i &,,              &28  &1.26   &&2.44 &&329.71    &&,,      && &6, 21    & 7.82    &7.40    &&PV & 5.57 & 5$^{h}$47$^{m}$& 17$^{h}$15$^{m}$\\
                                                  
04.10.2020&g &6$^{h}$20$^{m}$ &218 &2.29   &&1.53 &&5204.26   &&269.50  && &53,162   &37.95    &1.66    &&V  & 29.37& 1$^{h}$52$^{m}$& 5$^{h}$17$^{m}$\\
          &r &,,              &218 &3.78   &&1.53 &&22101.01  &&,,      && &53,162   &90.45    &1.66    &&V  & 21.54& 1$^{h}$14$^{m}$& NA\\
          &i &,,              &218 &4.83   &&1.53 &&11734.01  &&,,      && &53,162   & 4.33    &1.66    &&V  & 22.96& 2$^{h}$34$^{m}$& 3$^{h}$43$^{m}$\\
                                                  
05.10.2020&g &5$^{h}$50$^{m}$ &200 &7.32   &&1.53 &&14187.42  &&249.45  && &49,150   &153      &1.76    &&V  & 30.28& 3$^{h}$15$^{m}$& 33$^{h}$57$^{m}$\\
          &r &,,              &200 &41.49  &&1.53 &&91965.48  &&,,      && &49,150   &488.4    &1.76    &&V  & 23.16& 6$^{h}$37$^{m}$& NA\\
          &i &,,              &200 &1.92   &&1.53 &&62527.53  &&,,      && &49,150   &2.5      &1.76    &&V  & 30.14& 4$^{h}$17$^{m}$& 6$^{h}$41$^{m}$\\
                                                 
17.10.2020&i &6$^{h}$01$^{m}$ &199 &1.01   &&1.53 &&1247.19   &&248.33  && &49,150   &7.12     &1.76    &&V  & 12.92& 2$^{h}$44$^{m}$& 3$^{h}$56$^{m}$ \\
25.10.2020&i &0$^{h}$32$^{m}$ &22  &0.60   &&2.75 &&93.43     &&40.29   && &4 ,15    &1.53     &4.89    &&NV & ...  & ...&...\\
                                                  
31.10.2020&g &2$^{h}$40$^{m}$ &88  &0.60   &&1.84 &&263.74    &&121.77  && &21,66    &101.20   &2.20    &&PV & 19.02& 2$^{h}$20$^{m}$& 3$^{h}$48$^{m}$\\
          &r &,,              &93  &0.86   &&1.53 &&459.97    &&127.63  && &22,69    &5.57     &2.20    &&PV & 6.33 & 2$^{h}$33$^{m}$& NA\\
                                                  
01.11.2020&g &2$^{h}$50$^{m}$ &98  &1.51   &&1.53 &&510.55    &&133.48  && &23,72    &1.92     &2.12    &&NV & ...  & ...&...\\
          &r &,,              &100 &0.66   &&1.53 &&1143.73   &&135.81  && &24,75    &9.23     &2.12    &&PV & 8.32 & 1$^{h}$42$^{m}$& NA\\
          &i &,,              &99  &1.08   &&1.53 &&789.46    &&134.64  && &24,75    &9.86     &2.12    &&PV & 16.78& 1$^{h}$38$^{m}$& 2$^{h}$22$^{m}$\\
                                                  
05.11.2020&g &3$^{h}$07$^{m}$ &81  &0.63   &&1.84 &&327.27    &&113.51  && &19,60    &1.45     &2.20    &&NV & ...  & ...&...\\
          &r &,,              &99  &0.41   &&1.53 &&1694.38   &&134.64  && &24,75    &4.97     &2.12    &&PV & 12.06& 1$^{h}$20$^{m}$& 5$^{h}$50$^{m}$ \\
          &i &,,              &102 &0.96   &&1.53 &&675.10    &&138.13  && &24,75    &4.37     &2.12    &&PV & 26.74& 0$^{h}$41$^{m}$& 2$^{h}$10$^{m}$  \\
                                                  
20.11.2020&g &3$^{h}$10$^{m}$ &105 &1.35   &&1.53 &&401.84    &&141.62  && &25,78    &3.81     &2.12    &&PV & 46.54& 0$^{h}$36$^{m}$&52$^{m}$ \\
          &r &,,              &109 &0.39   &&1.53 &&2615.49   &&146.26  && &26,81    &6.13     &2.12    &&PV & 18.77& 1$^{h}$43$^{m}$& NA\\
          &i &,,              &112 &2.15   &&1.53 &&1229.17   &&149.73  && &27,84    &7.70     &2.03    &&V  & 46.46& 0$^{h}$55$^{m}$& 1$^{h}$18$^{m}$ \\
                                                  
23.11.2020&g &1$^{h}$42$^{m}$ &61  &0.96   &&1.84 &&205.09    &&89.59   && &14,45    &4.63     &2.52    &&PV &  3.85& 3$^{h}$46$^{m}$& 5$^{h}$26$^{m}$\\
          &r &,,              &66  &0.44   &&1.84 &&406.99    &&95.63   && &15,48    &0.10     &2.52    &&NV &  ... &...&...\\
          &i &,,              &62  &0.47   &&1.84 &&314.73    &&90.80   && &14,45    &3.98     &2.52    &&PV &  9.97& 3$^{h}$46$^{h}$& 5$^{h}$30$^{m}$\\

\end{longtable}
\end{sidewaystable*}

\begin{figure*}
\label{fig3}
\begin{minipage}{0.4cm}
\rotatebox{90}{{\large {Color Index}}}
\end{minipage}
\begin{minipage}{\dimexpr\linewidth-1.10cm\relax}
\hspace*{0.0cm}
\mbox{\subfloat{\includegraphics[scale=0.36]{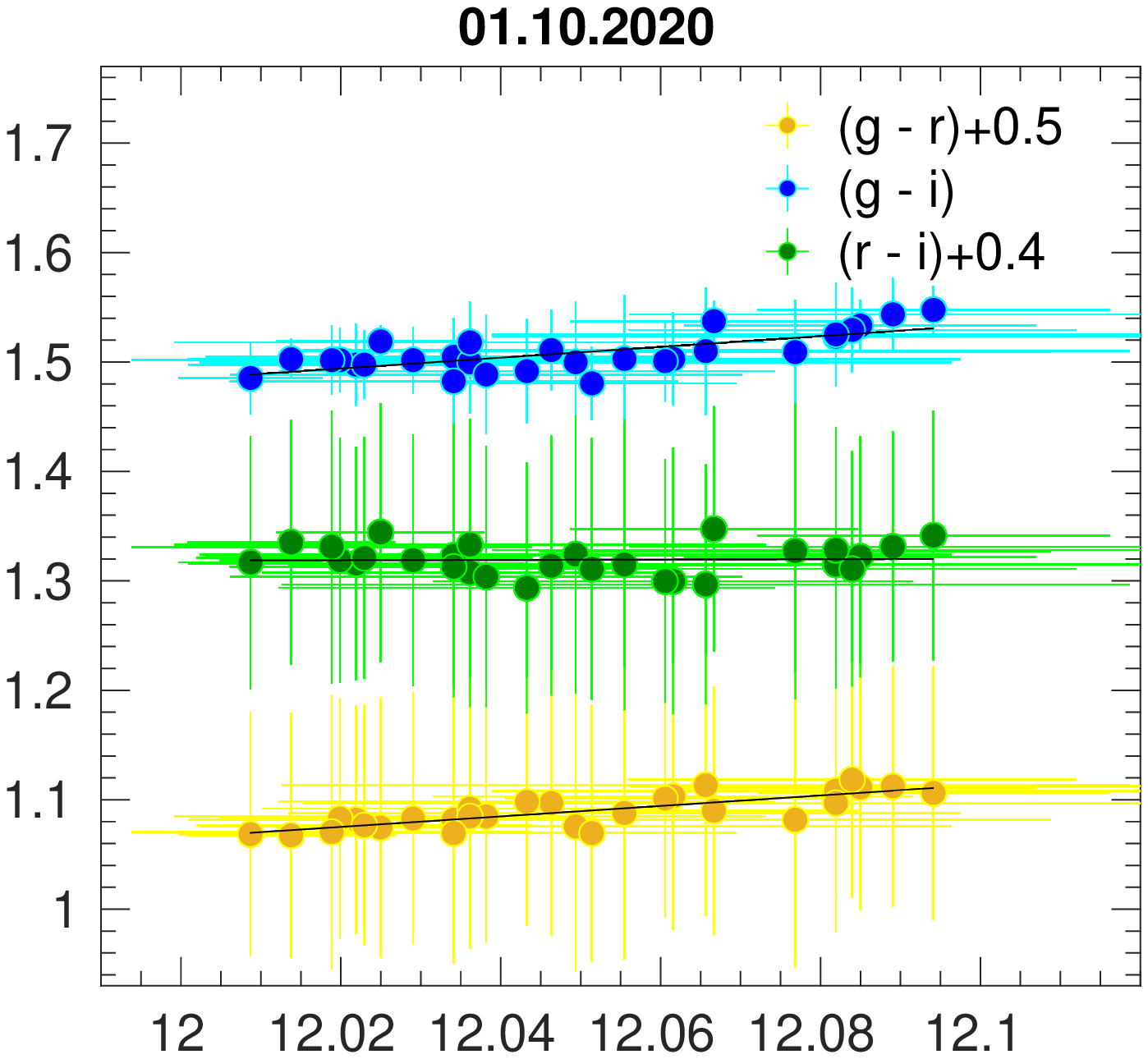} }\quad
\subfloat{\includegraphics[scale=0.36]{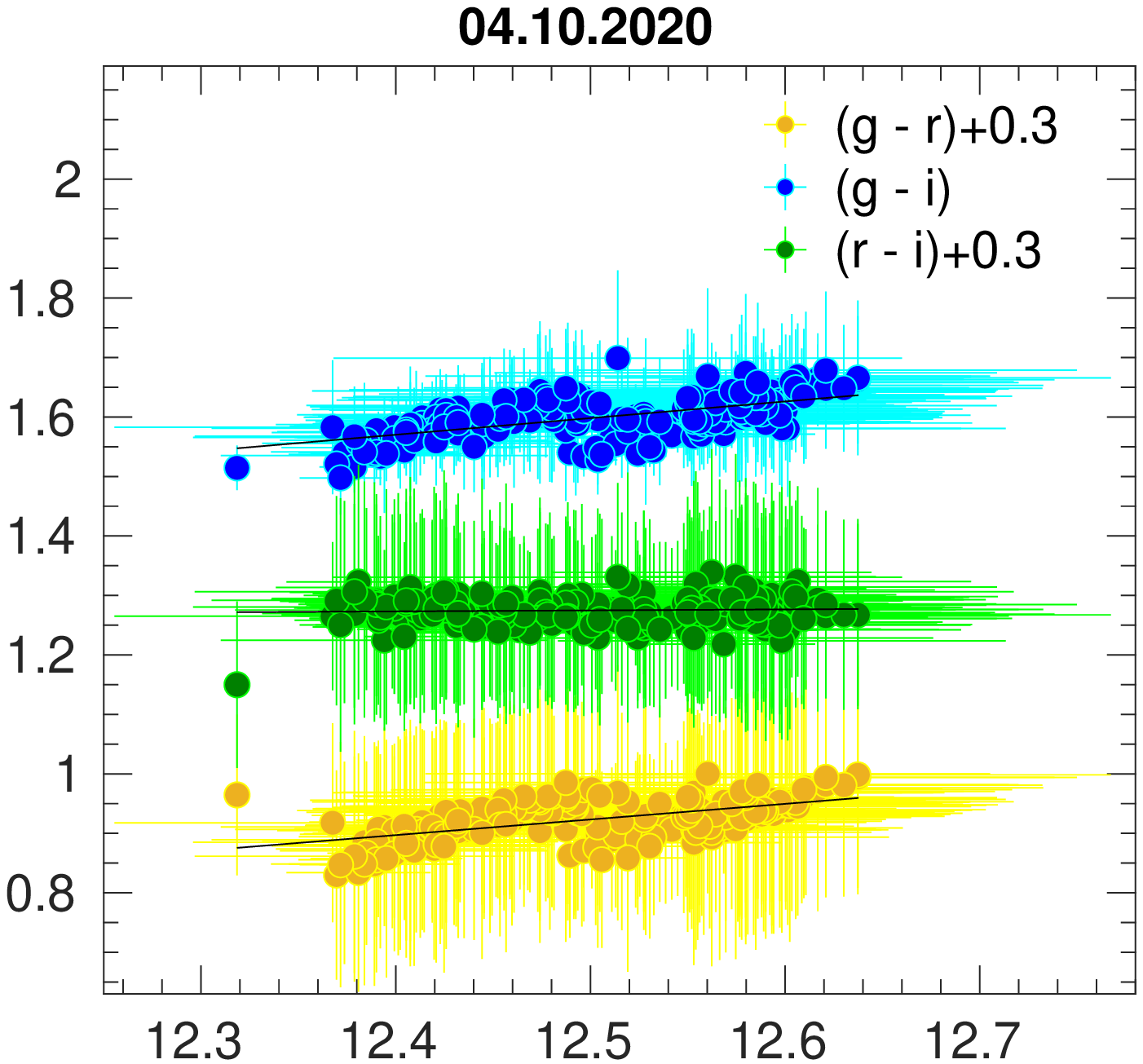} }\quad
\subfloat{\includegraphics[scale=0.36]{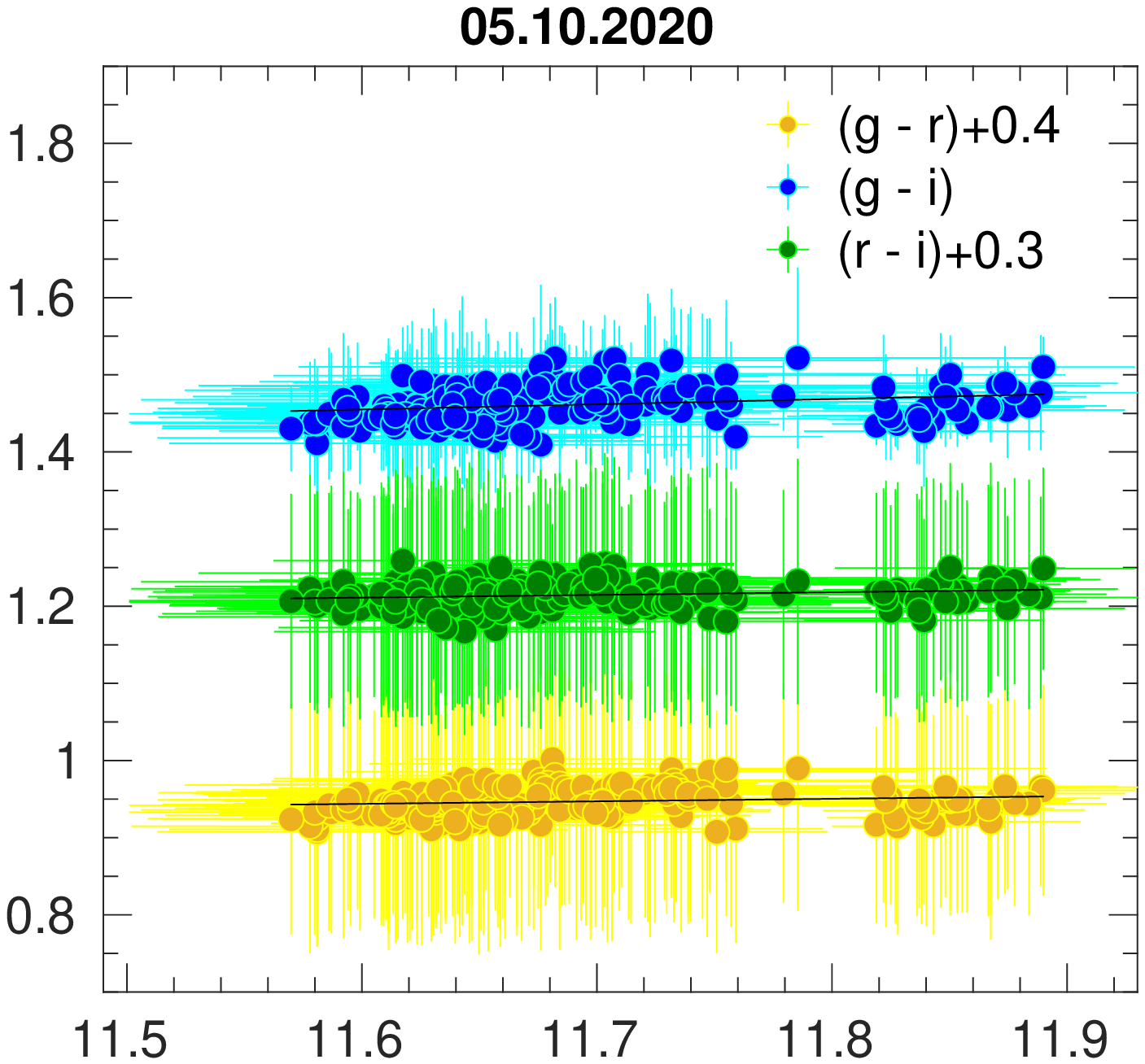} }}
\newline
\hspace*{0.1cm}
\mbox{\subfloat{\includegraphics[scale=0.36]{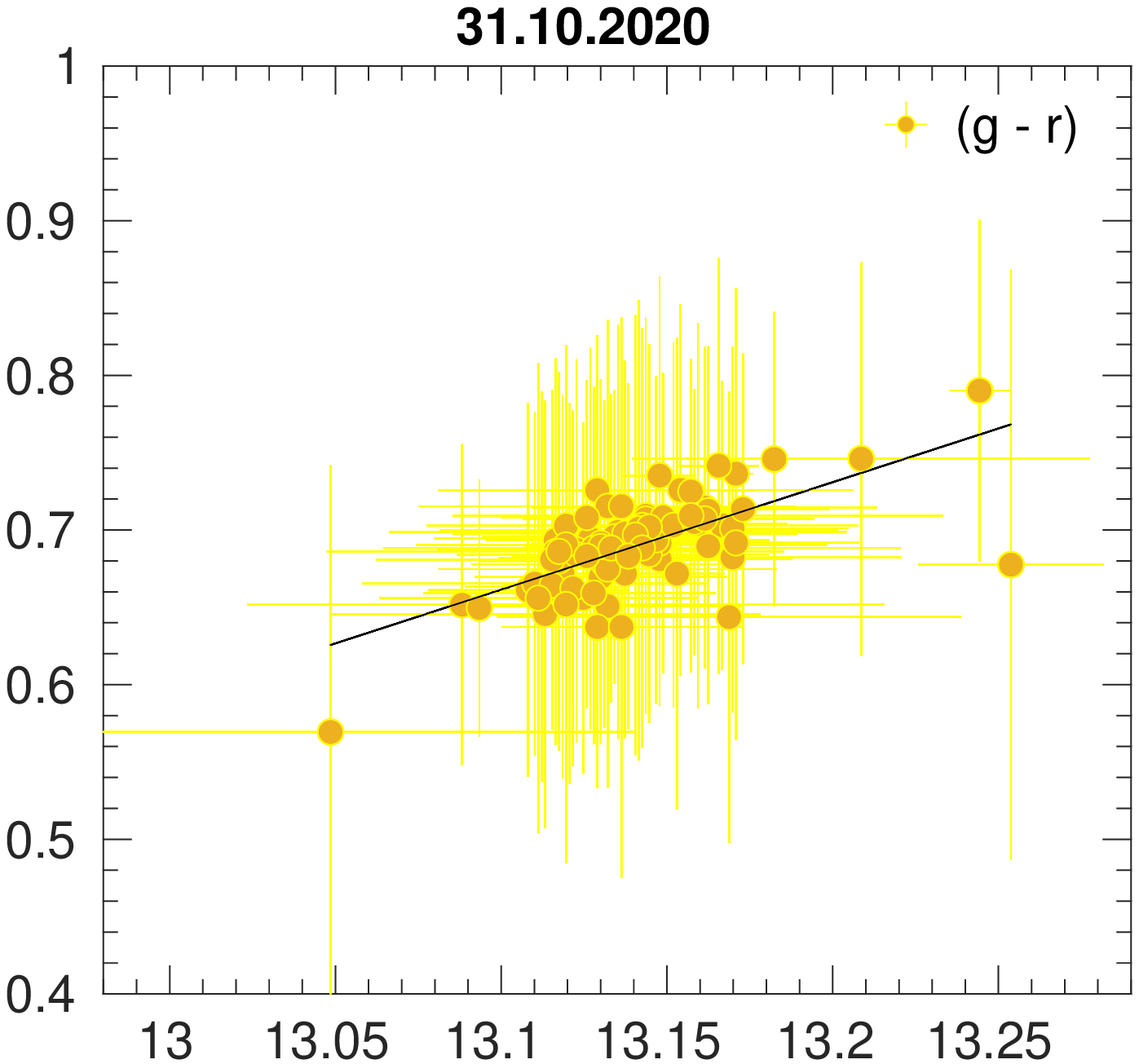} }\quad
\subfloat{\includegraphics[scale=0.36]{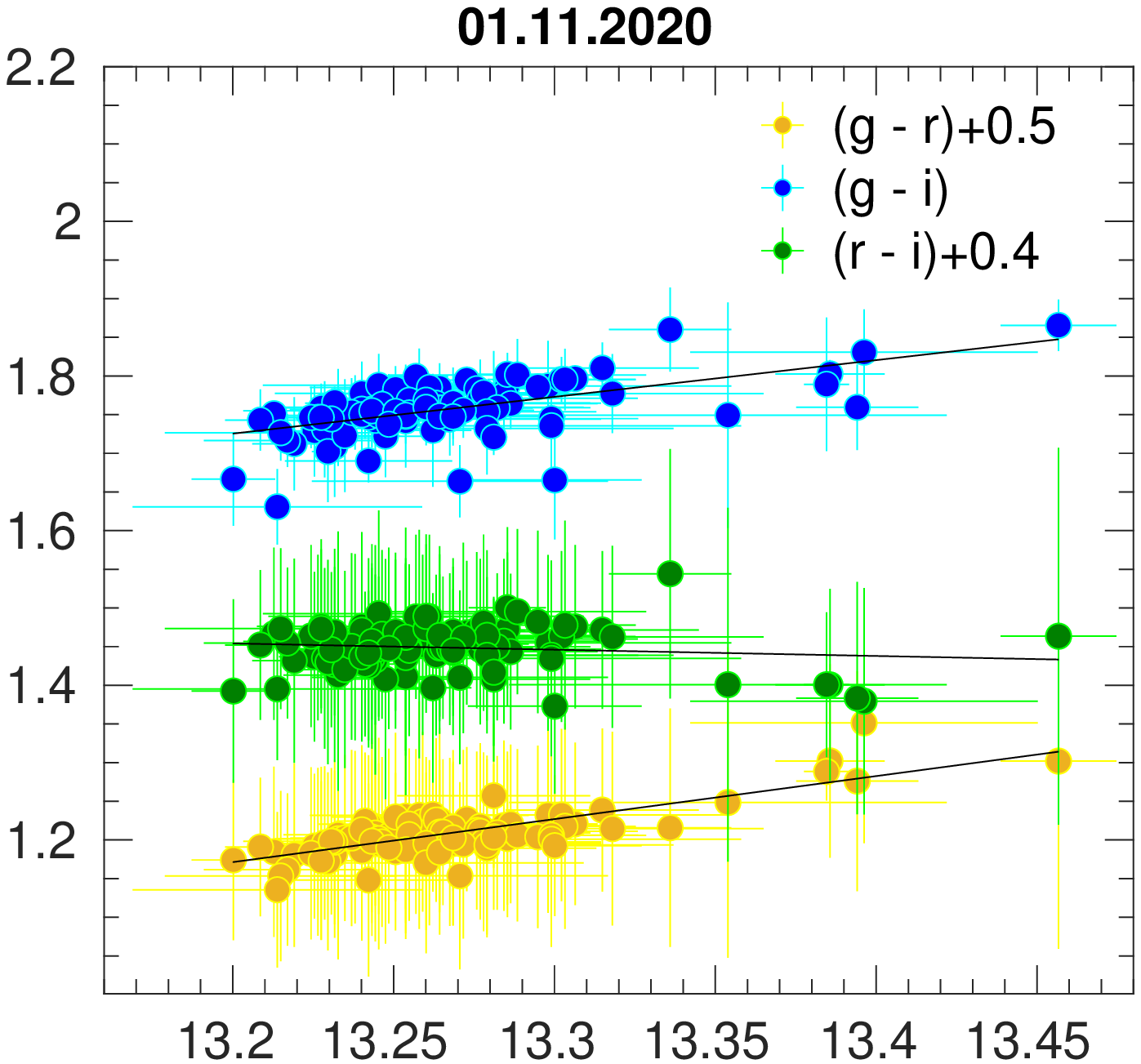} }\quad
\subfloat{\includegraphics[scale=0.36]{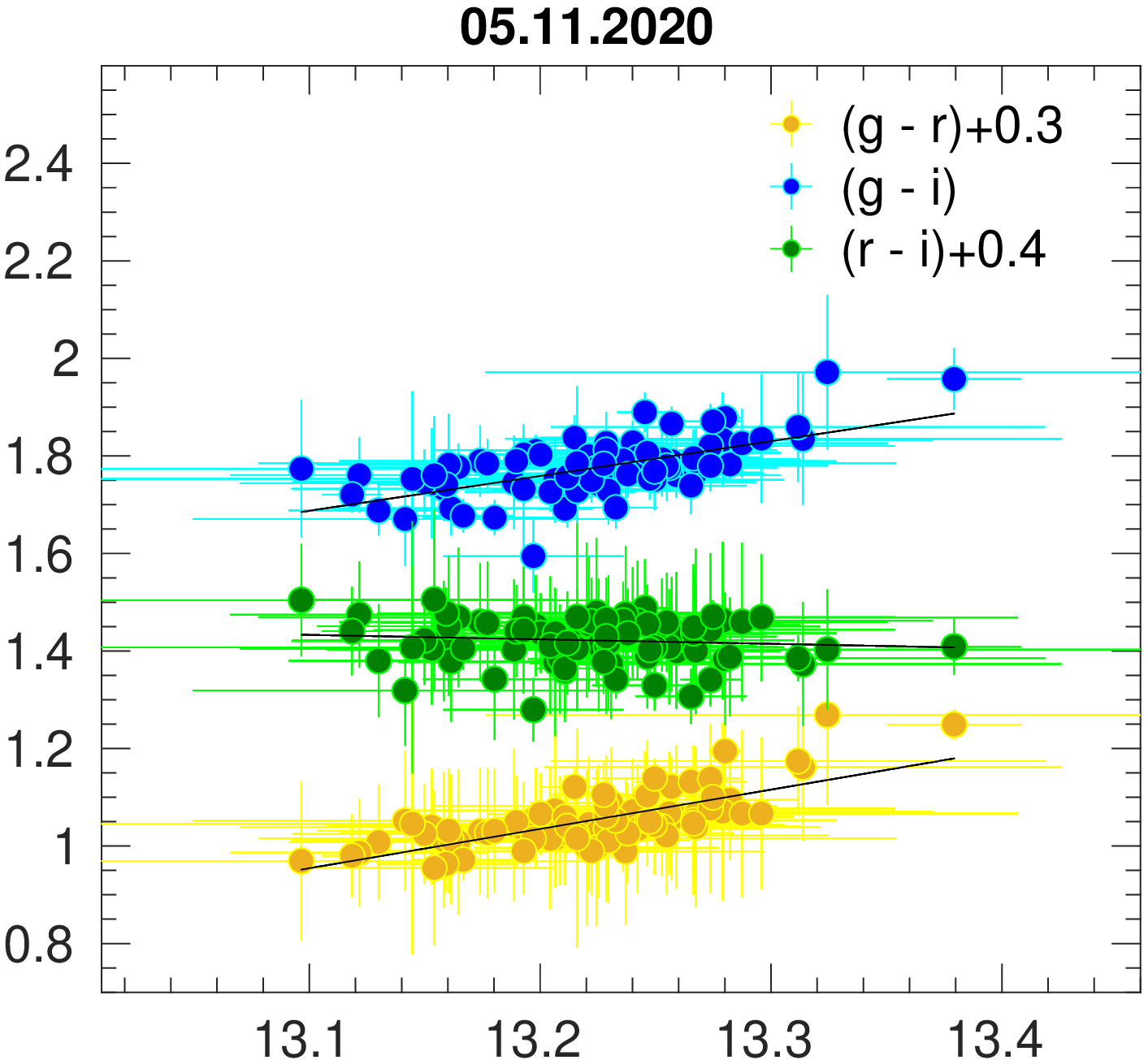} }}
\newline
\hspace*{2.8cm}
\mbox{\subfloat{\includegraphics[scale=0.36]{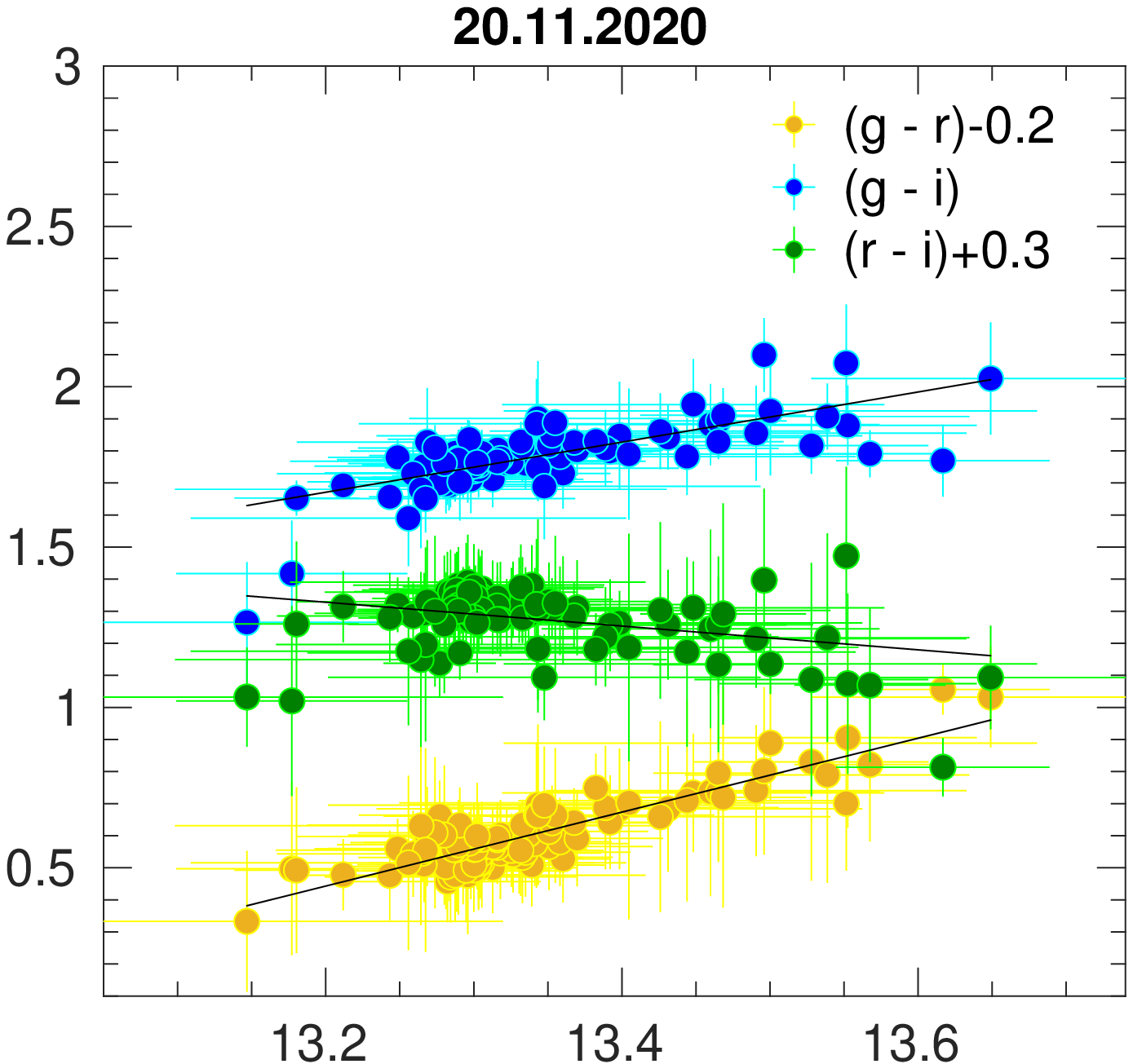} }\quad
\subfloat{\includegraphics[scale=0.36]{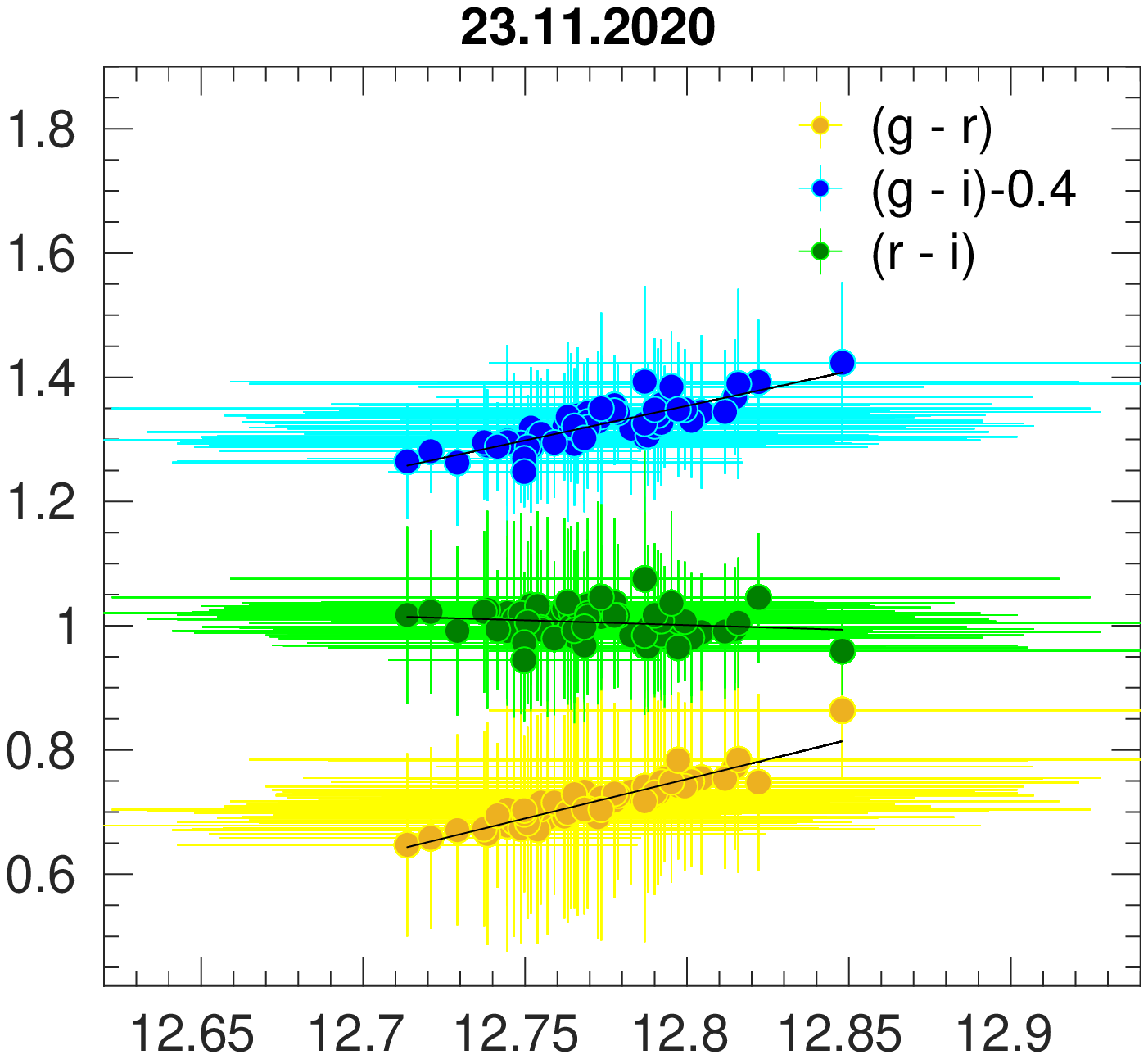} }}

\vspace*{0.3cm}\hspace*{7.0cm}{\large{g [mag]}}
\caption{Color indices vs. g-band magnitude plots on intranight timescale. Observing dates are given at the top of each plot. The black lines represent a least squared fitting to the data points. Magnitudes used here are corrected for Galactic reddening and host galaxy contamination.}
\end{minipage}
\end{figure*}

On LTV timescale, the source was significantly variable in all the three bands with variability amplitudes $160.11\%$, $137.02\%$, and $132.04\%$ in g, r, and i bands, respectively. Unlike on IDV timescales, the source showed clear frequency dependent variability in this case. The overall brightness variation can be illustrated with i-band magnitudes as it has most observational nights. In the beginning, on Oct. 1, the source brightness was $11.5$ mag which decreased to $11.9$ on Oct. 4 followed by brightening to $11.2$ mag on Oct. 5, the peak of overall LC in our observational session. A concave LC was detected on Oct. 5, in the way that the source brightness gradually increased from $11.4-11.1$ mag, then decreased to about 11.2 mag (see Figure 2). After that, the source stayed relatively stable at a faint state ($\sim 12.2 - 12.5$ mag), and finally, brightened to $\sim12$ mag at the last night of our observational session (see Figure 3). Overall in this period, the brightness changed by 1.3 mag from a minimum of 12.5 to a maximum of 11.2. On long timescale, the estimated $t_{v}$ values are 11.41, 11.72 and 11.97 days for g, r, and i-bands, respectively with or without including errors in eq. 6. 

In correlation analysis, we measured the DCFs values between the $g$ and the other two bands using a binning size of $5$ which is equivalent to a time difference of $\sim$ 7 minutes. We found that all the optical bands are nicely correlated to each other with DCF peaks at $\approx$ 0.80--0.95 i.e., with 80--95\% degree of correlation, having a DCF curve similar to that shown in Figure 4. In this case, the cross-correlation is very nearly symmetrical around zero lag, which is almost identical to an auto-correlation, explicitly indicating simultaneous emission of all the optical bands from a same electron population and radiation zone.

\subsection{Spectral Variation with Color--Magnitude Diagram}
In order to study the spectral behavior of the outburst event, we investigated the color--magnitude diagrams (hereafter CM diagram) of {\it g-r}, {\it g-i}, and {\it r-i} color indices vs. the {\it g} band magnitude. The magnitudes were corrected for Galactic extinction using the values taken from the NASA/IPAC Extragalactic Database\footnote{https://ned.ipac.caltech.edu/} \citep[NED,][]{1998ApJ...500..525S}. Since BL Lacertae is hosted by a relatively bright galaxy, we subtracted its contribution from the observed fluxes in order to avoid contamination in the color indexes. According to \citet{2000ApJ...532..740S}, the R band magnitude of the BL Lac host galaxy is R$_{host}$ = 15.55 $\pm$± 0.02 and by taking the average color indices for elliptical galaxies with M$_{V}<-21$ from \citet{2001MNRAS.326..745M}, we estimated the host galaxy brightness in other bands.  These Johnson filter band magnitudes are then converted into SDSS filter magnitudes following the steps given on SDSS website\footnote{https://www.sdss3.org/dr8/algorithms/sdssUBVRITransform.php} \citep{2005AJ....130..873J} and then corrected for the host reddening using the Galactic extinction coefficient values given on NED. These dereddened magnitudes are converted into SDSS fluxes using the method given on SDSS website. The resulting host galaxy fluxes in the g, r, and i bands are 2.62, 2.48, and 2.91 mJy, respectively. If we consider the source extraction radius \citep[following][]{2002A&A...390..407V, 2009A&A...507..769R} used in this study, the host galaxy contribution to the observed flux is about 50$\%$ per cent of the total galaxy flux. This contribution was removed from both the observed magnitudes and fluxes for color and spectral studies.

The CM-plots are  presented in Figure 5. As seen in the figure, most of the plots show a positive correlation characterized by hardening of the optical continuum with increasing brightness, a trend popularly known as bluer--when--brighter (BWB). Only in a few cases, an opposite trend is observed i.e., steepening of the continuum with brightness, also known as redder--when--brighter (RWB). To quantify these correlations, we performed a linear regression fit on each plot using the least-squared method. The fitting is shown by a black straight line in the plots. We considered it as significant only if the derived probability of rejecting the null-hypothesis, $p$-values $\geq 0.01$ (i.e 99\% significance level). This way, 72\% instances exhibit significant correlations and all of them follows a BWB trend. However, it is worth mentioning that 96\% of the light curves show a BWB trend, and the rest shows a very weak RWB pattern. All the fitting results are listed in Table 3. Overall, the {\it r-i} color correlations to {\it g--}band are weaker then the rest and those which show a weak RWB trends belong to this color index, but this pattern appear only in the later observation taken in November. We noticed that the color analysis is not much affected by the host galaxy contribution and Galactic extinction. 

In the CM diagram for 4th October (Figure 5), the night before the peak night we notice that the {\it g--r} and {\it g--i} color indices cluster in two different branches at a particular brightness ($\sim$ 12.49 mag), one below the other showing an overall BWB trend. For a clear representation, we plot only {\it g--i} data points against {\it g--}band magnitude in the 1st panel of Figure 6. We avoid the error bars for better visibility. In this figure, the two branches (red and yellow points) which belong to two different epoch clearly stand out. The branches represent two distinct spectral states which appear within only 6 hours 20 minutes of observing time. The second branch appears after $\sim$ 3 hours. The $R$ values obtained from separate fitting to the red and yellow branches are 0.82 and 0.74 with slopes 0.70 and 0.73, respectively. When we fit the the overall data, it gives a very weak correlation with a small R value as compared to the separate fitting values. The time evolution of color is a bit complicated rather than gradual as seen in the second panel of the figure. Although there is distinct branching of the colors, it seems the colors do not particularly follow a clean pattern with time within the individual branches. 

From the CM diagram, the average spectral indices of the optical spectrum can be derived simply by using the average colors $\langle g-r, g-i, r-i \rangle$  \citep{2015A&A...573A..69W}, as

\begin{equation}
\langle \alpha_{gr, gi, ri} \rangle = {0.4\, \langle g-r, g-i, r-i \rangle \over \log(\nu_{g, r} / \nu_{r, i})} \, ,
\label{eq_1}
\end{equation}

where $\nu_{g, r}$ and $\nu_{r, i}$  are effective frequencies of the respective bands \citep{1998A&A...333..231B}. The estimated values are listed in Table 3. 

The Galactic reddening and host corrected color indices vs. g-band magnitudes during the entire observing period is shown in the bottom panel of Figure 3. A least-squared fitting to the g-r, g-i and r-i colors gives slope of 0.13, 0.20, and 0.06 with correlation coefficient, R values 0.92, 0.98 and 0.79, respectively, which represents even stronger correlations than that is observed in the respective colors for intraday timescales. The corresponding spectral slopes estimated using eq. 7 are 3.39 $\pm$ 0.46, 3.47 $\pm$ 0.14, and 3.58 $\pm$ 0.63, respectively. From these we can say that the long term flux oscillations are as strongly chromatic as that in fast flux changes and follow a strong BWB trend.

\begin{table}
\small
\label{tab2}
\centering
\caption{\bf Results of color and SED analysis.}
\begin{tabular}{lcccccc} \hline \hline
Date  &Color  & R       & P         & a   &$\alpha_{gr, gi, ri}$& $\alpha_{sed}$   \\
      &       &(1)      & (2)       & (3) & (4) &  (5)                  \\\hline

01.10.2020 & g-r  & 0.80    & 2.46e-07 &  0.48    &3.10 $\pm$ 0.39& 0.92 \\
           & g-i  & 0.70    & 1.76e-05 &  0.49    &3.19 $\pm$ 0.08&     \\
           & r-i  & 0.03    & 8.96e-01 &  0.01    &3.33 $\pm$ 0.56&    \\

04.10.2020 & g-r  & 0.58    & 4.04e-21 &  0.26    &3.21 $\pm$ 0.59& 1.09 \\
           & g-i  & 0.60    & 7.50e-23 &  0.28    &3.37 $\pm$ 0.17&      \\
           & r-i  & 0.05    & 4.26e-01 &  0.02    &3.59 $\pm$ 0.77&      \\

05.10.2020 & g-r  & 0.15    & 3.67e-2  &  0.03    &2.96 $\pm$ 0.51& 0.82  \\
           & g-i  & 0.25    & 4.71e-4  &  0.07    &3.11 $\pm$ 0.14&       \\
           & r-i  & 0.18    & 1.19e-2  &  0.04    &3.31 $\pm$ 0.67&       \\

31.10.2020 & g-r  & 0.66    & 2.87e-12 &  0.70    &3.42 $\pm$ 0.47&  \\

01.11.2020 & g-r  & 0.78    & 6.49e-21 &  0.56    &3.48 $\pm$ 0.39& 1.40 \\
           & g-i  & 0.57    & 6.22e-10 &  0.48    &3.65 $\pm$ 0.09&  \\
           & r-i  &-0.12    & 2.28e-01 & -0.08    &3.90 $\pm$ 0.57&  \\

05.11.2020 & g-r  & 0.72    & 1.85e-14 &  0.80    &3.62 $\pm$ 0.41& 1.43  \\
           & g-i  & 0.62    & 4.29e-10 &  0.71    &3.69 $\pm$ 0.14&  \\
           & r-i  &-0.10    & 3.60e-01 & -0.09    &3.78 $\pm$ 0.56&  \\

20.11.2020 & g-r  & 0.89    & 2.59e-37 &  1.15    &3.78 $\pm$ 0.60& 1.44 \\
           & g-i  & 0.71    & 3.18e-17 &  0.78    &3.69 $\pm$ 0.19&  \\
           & r-i  &-0.35    & 2.12e-04 & -0.37    &3.56 $\pm$ 0.82&  \\

23.11.2020 & g-r  & 0.92    & 3.11e-26 &  1.27    &3.52 $\pm$ 0.52& 1.34 \\
           & g-i  & 0.84    & 1.62e-17 &  1.11    &3.60 $\pm$ 0.20&  \\
           & r-i  &-0.16    & 2.20e-01 & -0.16    &3.72 $\pm$ 0.60&  \\\hline
\end{tabular} \\
{\bf Columns:} (1) Pearson correlation coefficient; (2) Probability of rejecting the null-hypothesis; (3) Slope of least squared fitting; (4) average spectral indices of the optical spectrum; (5) Spectral slope estimated from SED fitting. \\
\end{table}

\begin{figure*}
\centering
\mbox{\subfloat{\includegraphics[scale=0.46]{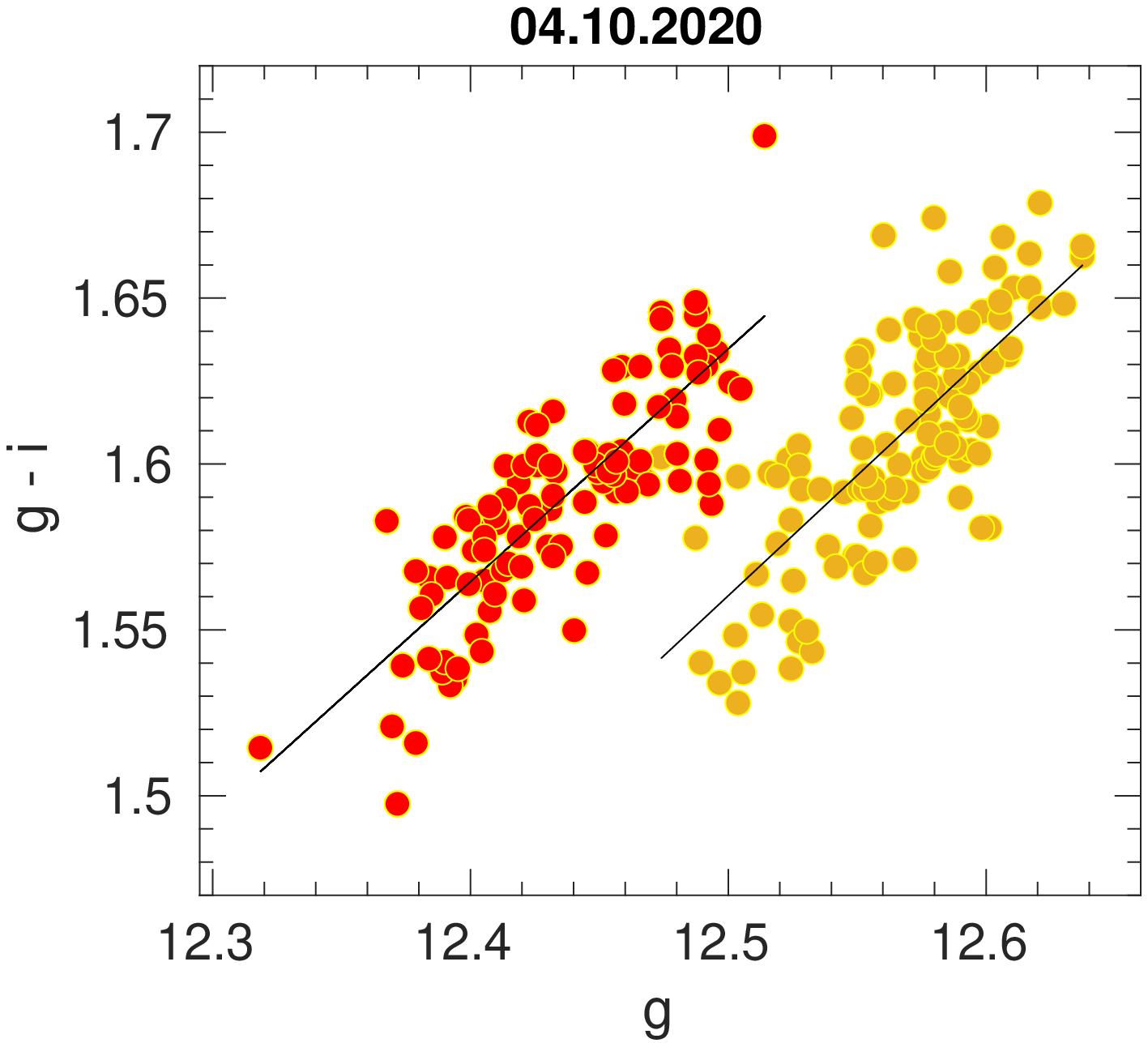} }\quad
\subfloat{\includegraphics[scale=0.45]{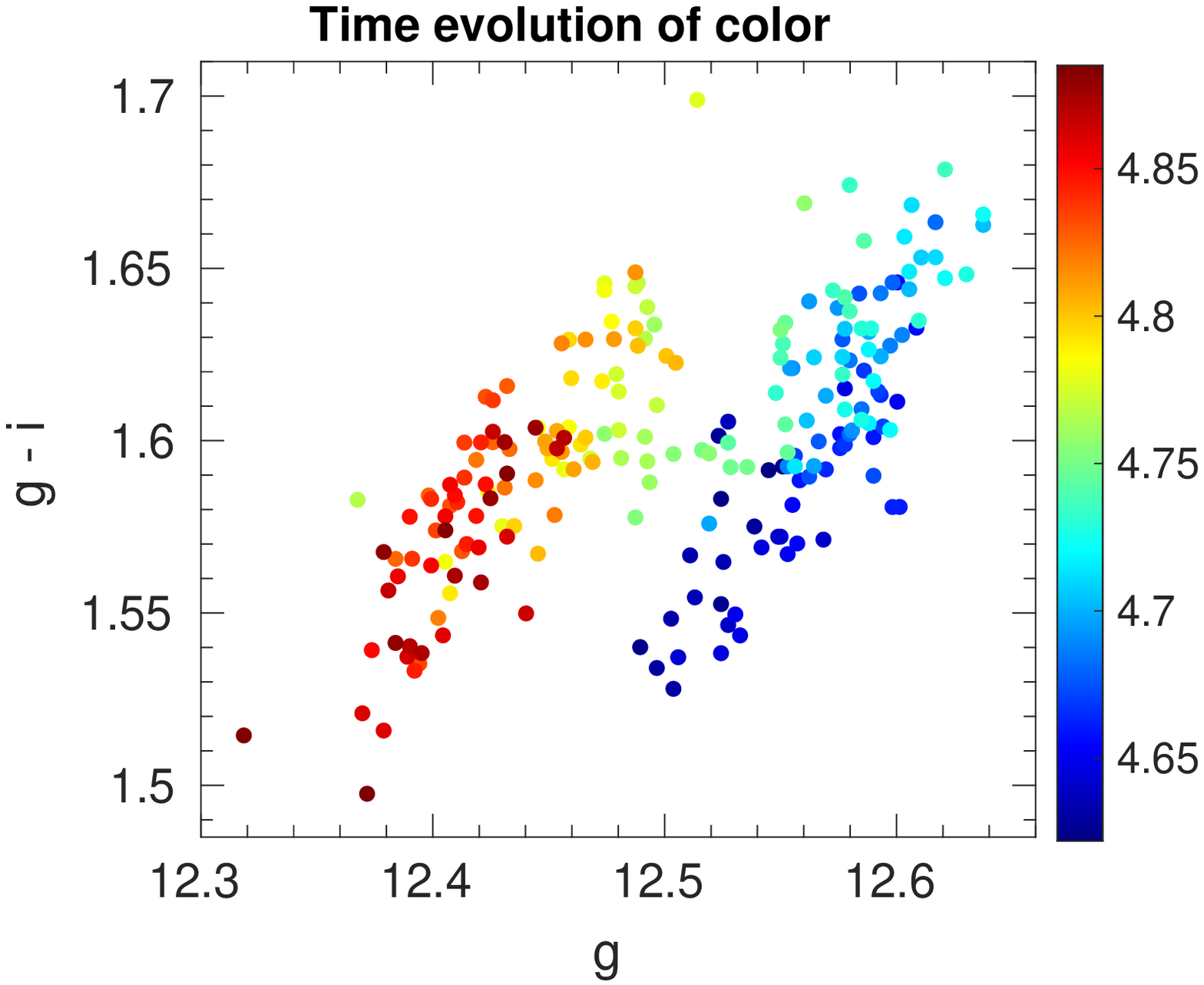} }}
\caption{We note two distinct optical spectral states appearing within $\sim$ 6 hours on October 4, 2020, the night before the source reached its first historical maxima in the optical band. A least squared fitting to the time separated data sets shown by the black lines gives Pearson's correlation coefficient 0.82 and 0.73, respectively shown by red and yellow symbols in the first panel. The error bars are omitted to get a more clear view of the spectral states. Second panel shows the time evolution of color, where the right hand side color bar represents time in the unit of + 2459122 MJD.}
\end{figure*}

\subsection{Spectral energy distributions}
Spectral energy distributions (SEDs) were developed with the gri bands using the average fluxes in individual energy bands for each nights. Fig. 7, shows the SEDs in the $log (\nu F_{\nu}$) vs. $log$ $\nu$ representation for 7 out of total 10 observing nights, for which data are available in all the three bands. The optical emission of blazars generally follows a power law shape of the form $F_{\nu}\propto \nu^{-\alpha}$, where $\alpha$ being the spectral index. Using this model, we extracted the optical spectral slopes for each nights which ranges from 0.82, corresponding to the hardest spectra observed on the night of outburst peak to 1.44, the softest spectra detected on Nov. 20. During our monitoring period, the slope varies by $\Delta\alpha = 0.62$ having a mean value of 1.2 where the bluest spectrum corresponds to the highest flux state of the source and the spectral hardness is strictly dependent on brightness. These values are listed in table 3.

In order to further explore how the optical emission evolve with time on the night of Oct. 4, we developed the gri-bands SEDs in a time interval of 15 minutes i.e., the data points in a given SED represent average of data acquired within 15 minutes of exposure. For SED fittings, we follow the same process explained above and extracted the spectral indices which ranges between 1--1.2. The variation of spectral indices throughout the night with time is shown in Fig. 8, where the vertical line represents the epoch where the double spectral state starts to appear in Fig 6. A polynomial of the form of $y = ax{^2} + bx + c$ can well represent the underlying spectral evolution trend of the night which is shown by the red curve in the figure. The model coefficients are $a =-8.6893$; $b =82.409$; and $c=-194.2485$.

\begin{figure}[]
\centering
\hspace{2.3cm}
\mbox{\subfloat{\includegraphics[scale=0.46]{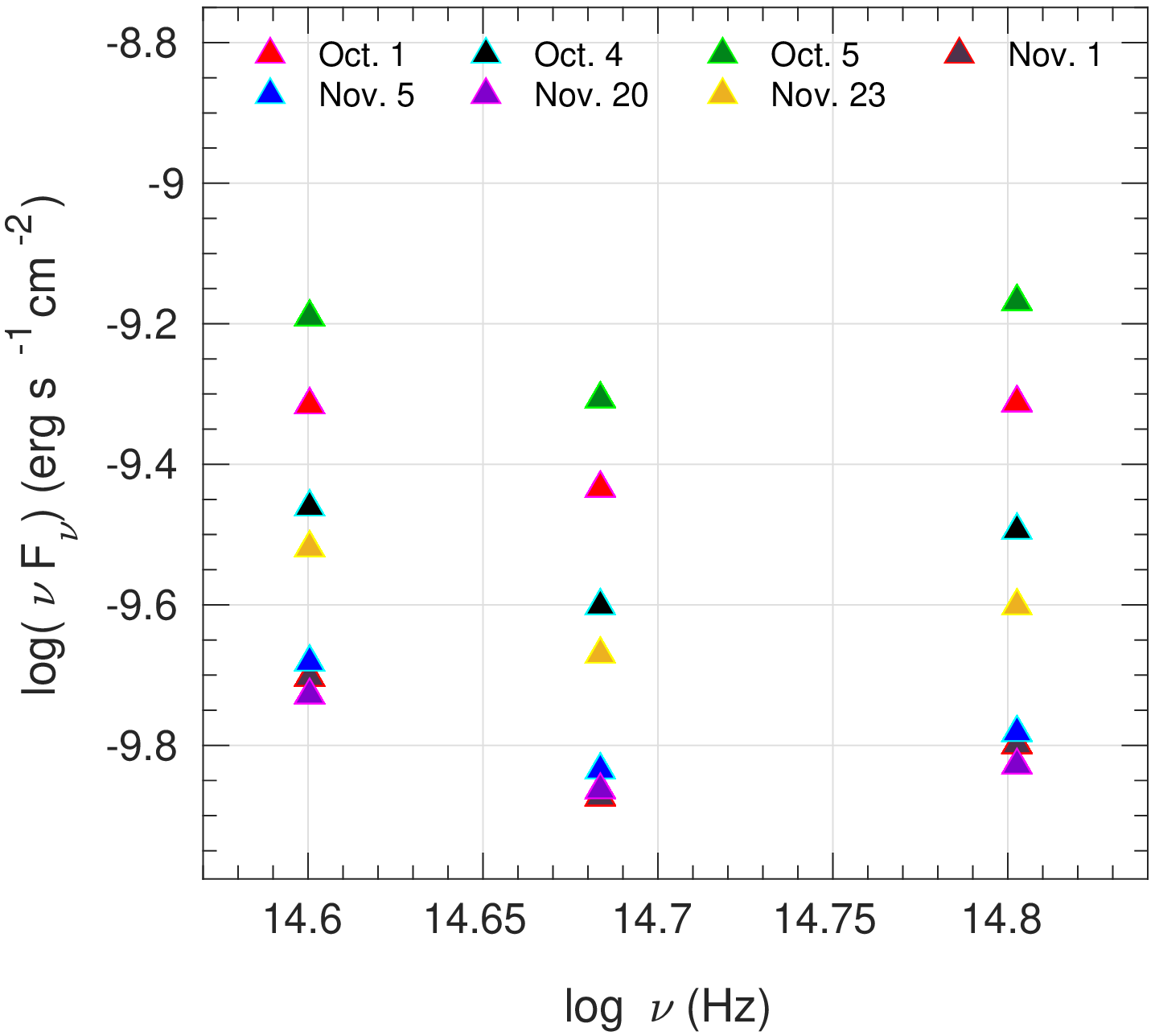} }}
\newline
\mbox{\subfloat{\includegraphics[scale=0.46]{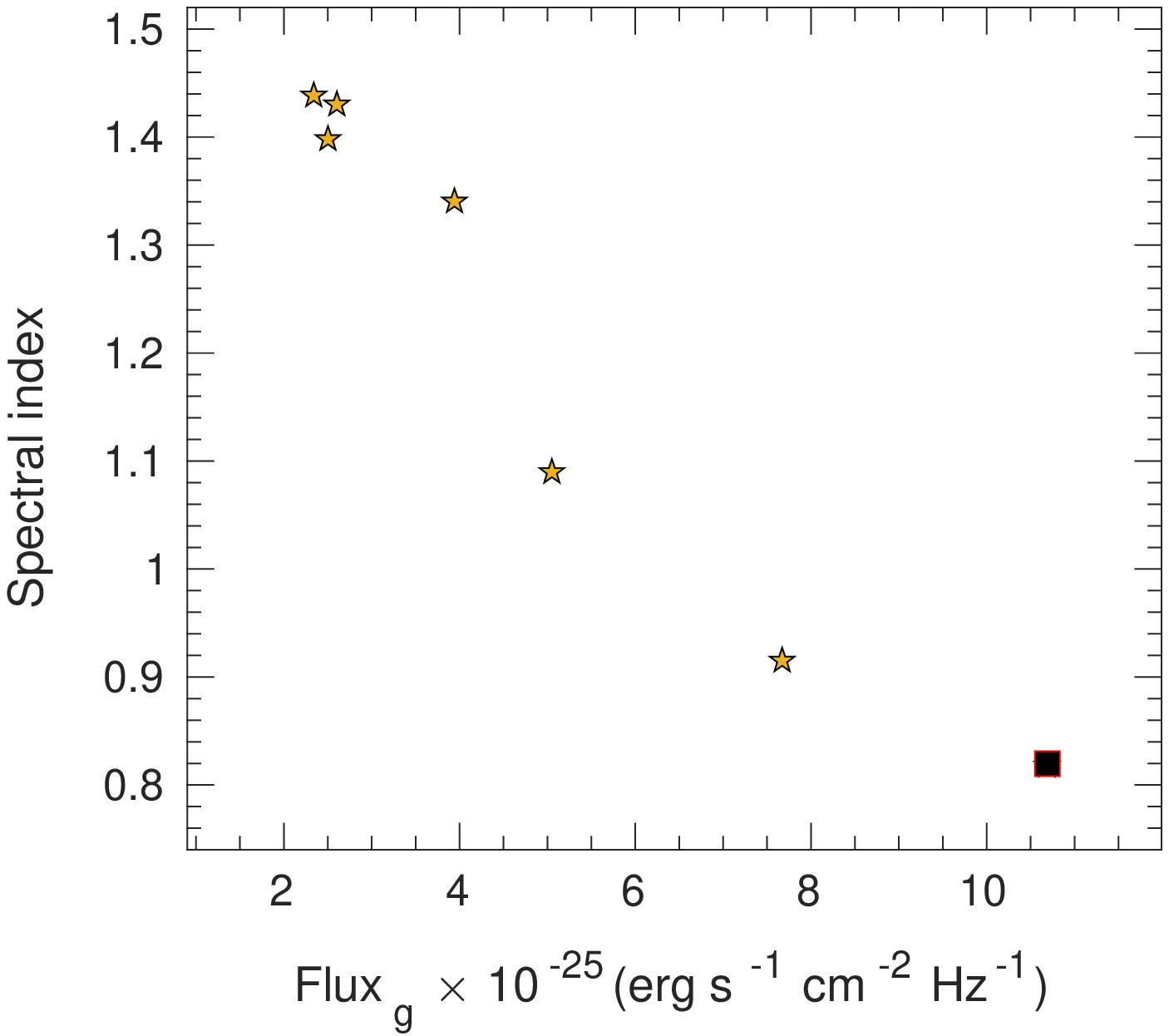} }}
\caption{Top panel: optical intranight SEDs of BL lac during our observing period. Dates are given on the top of the plot. Bottom panel: spectral slope and g-band flux relationship. The square point represent the outburst peak.}
\end{figure}

\begin{figure}[]
\centering
\includegraphics[scale=0.51]{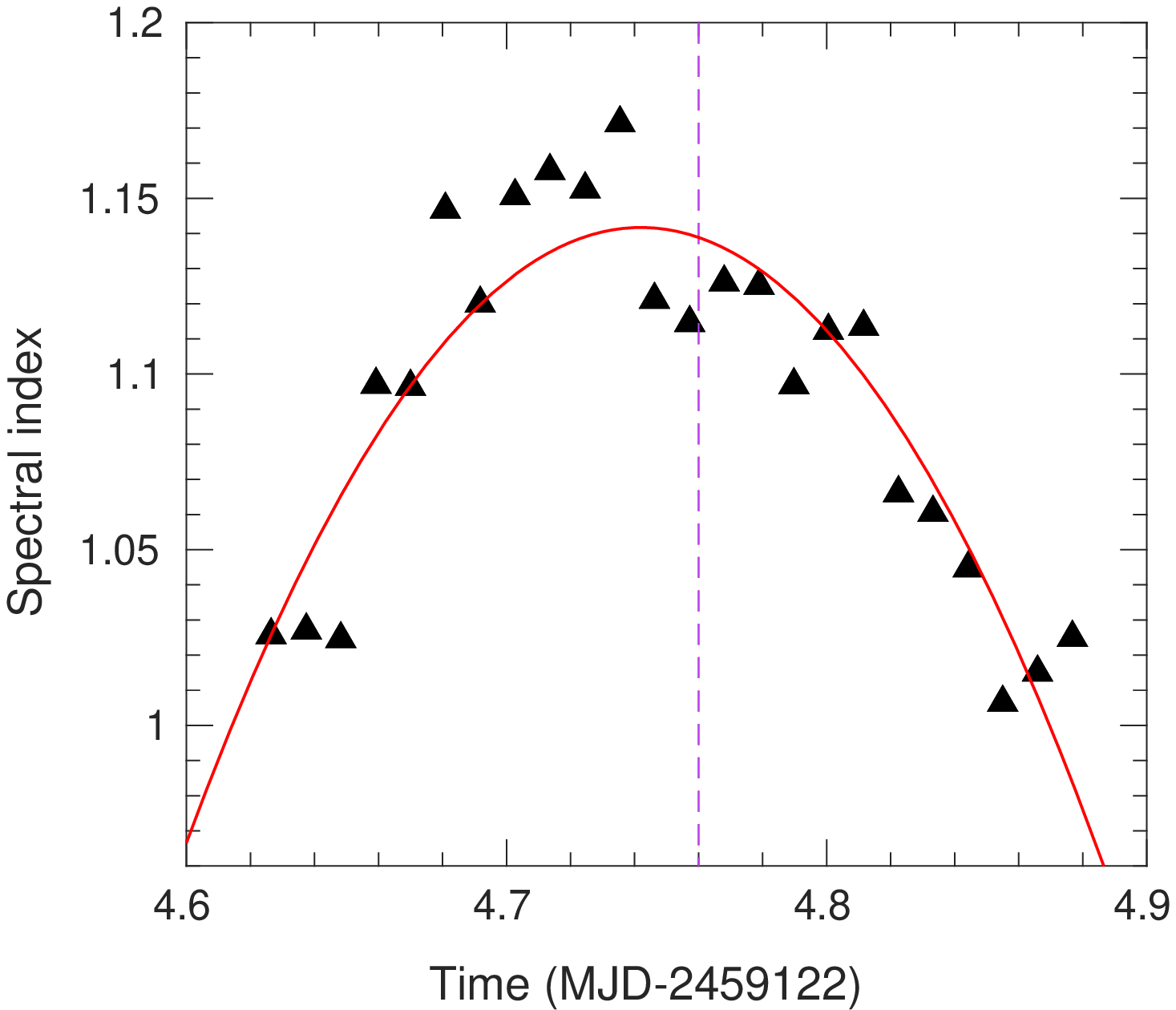}
\caption{ Spectral index evolution on Oct. 4. Slopes were estimated by fitting the SEDs with a simple power law model ($F_{\nu}\propto \nu^{-\alpha}$). The vertical dashed line represents the epoch where the double spectral states start to appear (see figure 6.). The spectral evolution through out the night can be fitted with a 2$^{nd}$ degree polynomial shown by the red curve.}
\end{figure}

\section{ Summary and Discussion}
Study of optical flares are extremely useful to get an in depth understanding on evolution of hidden physical processes that triggers or sustain them as they take longer timescales to develop than their corresponding higher energy counterparts, thus containing more structure for study. The historical high outburst of the prototype blazar BL Lac gives us an excellent window to examine such events in detail. In this paper, we carried out a study to understand the optical behavior of the blazar BL Lac during its outburst in 2020. We studied the flaring event on temporal and spectral domain on intraday and long timescales using multi-band optical observations from Oct. to Nov. 2020. 

From our time series analysis, we found that the source was significantly variable on both intranight and long timescales during the observing period with high amplitude of variation on long timescale than that on intranight. We found that the source showed variety of variability timescales on different nights from minutes to hours. The minimum timescale can be related to the size $l$ of the emitting feature in the jet (e.g. Romero et al. 1995b, 1999):

\begin{equation}
l \sim t_{v}c\gamma^{2}(1+z)^{-1}
\end{equation}

where $\gamma = (1-\beta^{2})^{-1/2}$ is the Lorentz factor of the shock and $z$ is the redshift. If we assume a small viewing angle of the jet which it is in case of blazar ($\cos\theta \sim \beta$), we can replace the Lorentz factor $\gamma$ in the equation by the Doppler factor $\delta = [\gamma(1-\beta \cos\theta)]^{-1}$. Assuming a typical value of $\delta \sim 10$, we get the emitting feature size $l \sim 4 \times 10^{-3}$ pc from the minimum timescales estimated from considering only variable instances. The highest variability timescale belong to a size of $\sim 6.7$ pc on IDV timescale. The compactness of the size of the emitting regions indicates that these variations are similar to light crossing time in $\sim 100$ Schwarzschild radius of the central black hole \citep[$\sim 2 \times 10^{8}\, M_{\odot}$;][]{2002ApJ...579..530W} which is coming from some mini-structures within the jet.

In the context of shock-in-jet model, the size or thickness of the emitting zone increases with the distance traveled by the shock along the jet \citep{1976PhFl...19.1130B}. An emitting region in the order $\times 10^{-3}$ pc estimated above means that the shock-feature interaction must be occurring very close to the jet’s apex. The asymmetric flux variation on nightly timescale observed in BL Lac in this study most likely resulted by some random processes, such as stochastic acceleration process in a turbulent region behind the shock front.
During the event, timescale of variability was even faster ($\sim 1^{m}$) in the X-ray band suggesting an even smaller size for the emitting region \citep{2022MNRAS.509...52D}. When QPOs in optical flux and polarization, and $\gamma$-ray flux with cycles of 13h were also detected \citep{2022Natur.609..265J}. 
 
On longer timescale, taking the average values of $t_{v}$ , we found an emitting region of size $\sim 55.07$ pc which produce. Unlike the IDV, the LTV shows frequency dependent behavior. Also, the variability timescales for the g, r, and i bands are systematic while we found chaotic timescales for the 3 energy bands in the same night on intranight timescales. These opposite behaviors clearly indicates that the emission on the two timescales are governed by completely different processes. In previous studies, different emission processes resulting the faster and longer timescale variability were reported for BL Lac where the chromatic faster variability and the mildly chromatic or achromatic longer variability have been linked with substructure in a shock induced jet and changes in the Doppler factor, respectively \citep{2002A&A...390..407V, 2004A&A...424..497V, 2009A&A...507..769R, 2013MNRAS.436.1530R}. Our results are similar with the previous one on intranight timescales, however, unlike on previous occasions, this time the source showed a strong BWB trend or chromatism on longer timescales as well. Considering the symmetric variability timescales for LTV and pc scale size ($\sim$ 55 pc) of the emitting region, we can say that the LT variation is likely originated from external jet region via a systematic process.

Absence of time lags between intra-optical bands is expected as they are very close in energy domain, and in case any lags are present, they must be insignificant and shorter than the resolution of the LCs. Simultaneous and correlated emission of the optical band and $\gamma$-rays were also detected during the event, indicating cospatiality of the emitting regions \citep{2022Natur.609..265J}. However, BL Lac was found to show correlated variability with hard lag of approximately 1 day between the X-rays and other energy bands in a study made by \citet{2021MNRAS.507.5602P}. In a shock-induced flare, the higher energy (HE) emission peaks first, which are followed by optical, IR and then radio, called as soft-lag which are commonly observed in blazars, and are explained by cooling of HE particles gradually radiating on lower and lower frequencies. However, an opposite pattern called as hard lags i.e., the HE photons lag behind the soft, has also been observed in a few well known HBLs; Mrk 401, Mrk 501 and PKS 2155$-$304 \citep{2002ApJ...572..762Z, 2005A&A...443..397B, 2009MNRAS.393.1063T, 2017ApJ...834....2A}. This type of observed delay requires the emitting particles within the jet to get heated or accelerated during the flare instead of getting cooled off via radiation \citep{2008A&A...491L..37M}. One possible reason could be generation of turbulence, especially the non-linear one by the particles themselves which could be efficient to trigger fast acceleration that leads to hard lags \citep{2009MNRAS.393.1063T}. In the context of particle acceleration within a shocked region, soft lags are expected when cooling timescales ($t_{cool}$) $\gg$ acceleration timescales ($t_{acc}$) of the relativistic particles and we observe hard lag when $t_{cool}$ $\sim$ $t_{acc}$ \citep[in depth discussion is given in][]{2002ApJ...572..762Z}. The particles could be accelerated to relativistic speed by the first-order Fermi acceleration in presence of an ordered magnetic field at a shock front and/or by the second-order Fermi acceleration, also known as stochastic acceleration process due plasma turbulence resulted by an chaotic magnetic field in a downstream region of the shock, and/or by multiple magnetic reconnections \citep{2006A&A...453...47K}. 

Another potential explanation for the observed variability behavior is magnetic-reconnection inside a magnetically dominated jet \citep{2011ApJ...726...62M, 2015MNRAS.450..183S, 2016MNRAS.462.3325P}. When the traveling shock interacts with the inhomogeneous medium, turbulence would be created behind the shock front by hydrodynamical instability \citep[e.g.,][]{2011ApJ...726...62M, 2014MNRAS.439.3490M}. The turbulence would locally amplify the magnetic field as filamentary structures.
A turbulent plasma with fast-moving magnetic filaments is likely a site for second-order Fermi acceleration of charged particles. Magnetic-reconnection events are expected to take place as the turbulent magnetic field behind the shock front would become progressively stronger due to continuous interaction of the traveling shock with its inhomogeneous medium. A strong magnetic reconnection could produce mini-jets, that is similar to the jet-in-jet model \citep{2009MNRAS.395L..29G}. In this scenario, large amount of magnetic energy get released whenever opposite polarity magnetic field lines interact with each other, which energized the intervening plasma and thus, accelerates the particles resulting the observed asystematic fast variability and hard lags. During the outburst of BL Lac, a kink in the jet at 43 GHz was observed by \citet{2022Natur.609..265J}, which yields presence of a tight helical magnetic field. The authors described the detected QPOs as a result of current-driven kink instabilities near a recollimation shock that was produced from pressure mismatches between the jet and it's surrounding. The systematic long term variability pattern on parsec scale that we observed in our study is most likely resulted by this helical magnetic field.

The spectral evolution during the outburst phase shows strong BWB chromatism which is observed in the entire observing period, except four cases of {\it r - i} color where a weak opposite pattern appeared in the later part of our observations. Absence of the RWB in other color on the same nights make us uncertain about these results. The color index and magnitude relation is governed by contributions made by the jet and the disk towards the overall optical emissions \citep{2009A&A...507..769R, 2012ApJ...756...13B}. The accretion disk component is inherently bluer and stable, while the jet contribution is variable and redder. The observed BWB states of the source is likely a result of significantly less radiative cooling of highly accelerated electrons over an under-luminous accretion disk. An interesting finding of this work is the detection of separate optical states on a single night observation. On October 4, the color indices corresponding to two epoch separated by $\sim$ 3 hours cluster in two distinct branches on the CM diagram both following BWB trend with similar slopes at high correlation significance, however the color of two branches are clearly different with bright branch being systematically redder than the faint one. This is the shortest period ever detected which exhibit such pattern. A weak similar type of detection was reported for the blazar PKS 2155--304 on years long period, while the source was transitioning from a flaring to a quiescent state \citep{2014A&A...571A..39H}. The authors related the observed pattern to distinct $\gamma$-ray states, where complex scaling between the optical and  $\gamma$-ray emission exist. In the study they found that the $\gamma$-ray flux depends on a combination of optical flux and color rather than flux alone. 

While both results were found at high optical flux state, our finding differs from theirs at two folds. Firstly, the time scale is completely different. Two tracks in this work were found in very short period in about six hours, while theirs were based on the comparison with the well-separated months-long variations in years. Secondly, our two tracks transited at $g\sim13.55$, around which the source brightness is gradually changing from $g\sim 13.6$ to $g\sim 13.5$ (see Figure 2), and there are no much overlapping flux. In contrast, the two color trends in PKS 2155-304 have large portion of overlapped flux states (see their Fig. 1). It seems that in our source, a physically distinct event happened when the source gradually brightened to $g\sim13.55$. It is likely that we were witnessing a new jet ejection event in the night of Oct. 4, 2020. While the ejection increases the source brightness, due to some reason, the new ejection has softer electron energy distribution than the previous one, which results in two tracks in CMD although both of them follow BWB trend.

The spectral evolution pattern we see from the time resolved spectral fit in figure 8 follows a hard-soft-hard (HSH) trend, that is the spectra evolve from hard to soft and then back to hard state again. Such HSH trend was found in Solar flares in X-rays at higher energies, which are suggested as general trend of non-thermal emissions throughout flares that are concerned with individual peaks in emission within a single flare \citep{2009ApJ...694L.162S}. In this scenario, particle trapping, either by stochastic acceleration or wave scattering in MHD turbulent regions leads to enhanced high-energy emission versus low-energy emission in between acceleration episodes \citep{2002paks.book.....A}. Hence, the HSH-pattern spectra may be associated with multiple injections of nonthermal electrons. Any small sub-peak may denote a new injection to soften the original spectra, and the spectra are hardened again afterward (Melnikov $\&$ Magun 1998). To check this scenario and it's applicability in blazar flares however, requires more sensitive and statistically strong data than we have in our hand, as well as further elaborated studies.

\acknowledgments
\noindent
We thank the anonymous referee for comments and suggestions that made the manuscript more comprehensive. I would like to thank C.M. Raiteri for the valuable discussion. This work by Nibedita Kalita is partially supported by the Chinese Academy of Sciences (CAS) President’s International Fellowship Initiative (PIFI) Grant, No. 2020PM0029. NK, PJ, HZ, and XP acknowledge support from the Natural Science Foundation of Shanghai (20ZR1463400, 20ZR1473600, 21ZR1469800). MFG acknowledges support from the National Science Foundation of China (NSFC) (grant 11873073), Shanghai Pilot Program for Basic Research -- CAS, Shanghai Branch (JCYJ-SHFY-2021-013), and the science research grants from the China Manned Space Project with NO. CMSCSST-2021-A06. JHF acknowledges support from NSFC (U2031201 and 11733001) and Guangdong Major Project of Basic and Applied Basic Research (grant No. 2019B030302001). ACG is partially supported by CAS PIFI (grant no. 2016VMB073). AAS and RSB acknowledge support from the Bulgarian National Science Fund of the Ministry of Education and Science under grants KP-06-H28/3 (2018), KP-06-H38/4 (2019), KP-06-KITAJ/2 (2020) and KP-06-PN68/1(2022)


\end{document}